\documentclass{aastex62}
\turnoffeditone

\received{ddmmyy}
\revised{ddmmyy}
\accepted{ddmmyy}

\submitjournal{ApJ}

\shorttitle{}
\shortauthors{Duan \& Wang}

\begin{document}

\title{ORIGIN FOR THE PROMPT SPECTRAL EVOLUTION CHARACTERISTICS AND HIGH ENERGY EMISSION DURING THE X-RAY FLARE IN GRB 180720B}

\correspondingauthor{Xiang-Gao, Wang}
\email{wangxg@gxu.edu.cn}

\author[0000-0001-5487-4537]{Ming-Ya Duan}
\affil{GXU-NAOC Center for Astrophysics and Space Sciences,Department of Physics,Guangxi University,Nanning 530004,China \\}
\affiliation{Guangxi Key Laboratory for the Relativistic Astrophysics,Nanning 530004,China}

\author[0000-0001-8411-8011]{Xiang-Gao Wang}
\affil{GXU-NAOC Center for Astrophysics and Space Sciences,Department of Physics,Guangxi University,Nanning 530004,China \\}
\affiliation{Guangxi Key Laboratory for the Relativistic Astrophysics,Nanning 530004,China}

\begin{abstract}
The gamma-ray burst GRB 180720B is very peculiar. On one hand, some
interesting features have been found by performing the detailed time-resolved spectral analysis in the prompt phase. First, the `flux-tracking' pattern is exhibited both for the low energy spectral index $\alpha$ and the peak energy $E_{p}$ in the Band function. Second, some parameter relations show strong monotonous positive correlations, include $E_{p}-F$, $\alpha-F$, $E_{p}-\alpha$, and $E_{p}-L_{\gamma,iso}$ for all time-resolved spectra. Lastly, it should be noted that the values of $\alpha$ do not exceed the synchrotron limits (from $-\frac{3}{2}$ to $-\frac{2}{3}$). On the other hand, the photons with the energy of $\gtrsim$ 100 MeV were detected by LAT both in the prompt phase and afterglow. Notably, the 5 GeV photon was observed at 142 s after the GBM trigger. The spectrum of this burst in the LAT range can be described as $F_{\nu}\propto \nu^{-1.3} t^{-1.54\pm0.02}$ in the afterglow phase. And there are six GeV photons during the X-ray flare when the lower energy emission is fading to a weaker level. We try to give reasonable interpretations of the mechanism for prompt emission and the high energy emission (100 MeV to GeV) in the afterglow. The interpretations suggesting that synchrotron origin can account for the prompt emission and synchrotron self-Compton radiation can account for both the spectrum and temporal behavior of the 100 MeV to GeV afterglow emission that have been accepted by us.

\end{abstract}


\section{Introduction} \label{sec:intro}
Gamma-ray bursts (GRBs) are the brightest explosions in the universe. It's generally believed that they are from the black holes or magnetars since the death of massive stars or the mergers of compact binaries (BH-NS or NS-NS) \citep{1974ApJ...187..333C,1986ApJ...308L..43P,1989Natur.340..126E,1992ApJ...395L..83N,1993ApJ...405..273W,1999ApJ...524..262M,2006ARA&A..44..507W,2015PhR...561....1K}.  The observed gamma-ray burst spectra can be fitted well by a function named Band \citep{1993ApJ...413..281B} both for the time-integrated spectra and the time-resolved spectra. It is pointed out that the spectral parameters, such as low-energy \edit1{\replaced{powerlaw}{power law}} index $\alpha$ and peak energy $E_{p}$, are evolves with time instead of remaining constant. In the early years, \citet{1983Natur.306..451G}, \citet{1986ApJ...301..213N}, \citet{1994ApJ...422..260K}, \citet{1994ApJ...426..604B}, \citet{1995ApJ...439..307F}, \citet{1997ApJ...479L..39C}, \citet{2006ApJS..166..298K} \citet{2009ApJ...698..417P} in \edit1{\added{the}} pre-$Fermi$ era and \citet{2012ApJ...756..112L}, \citet{2016A&A...588A.135Y}, \citet{2018MNRAS.475.1708A}, \citet{2018arXiv181003129L}, \citet{2018arXiv181007313Y} in \edit1{\added{the}} $Fermi$ era have shown the evolution characteristics of $\alpha$ and $E_{p}$. And the evolution patterns have been summarised as three types in \edit1{\added{the}} pre-$Fermi$ era: (i) they are decreasing monotonically, named `hard-to-soft' trend \citep{1986ApJ...301..213N,1994ApJ...426..604B,1997ApJ...486..928B}; (ii) they are increasing/decreasing when the flux is increasing/decreasing, named `flux-tracking' trend \citep{1983Natur.306..451G,1999ApJ...512..693R}; (iii) `soft-to-hard' trend or chaotic evolutions \citep{1985ApJ...290..728L,1994ApJ...422..260K}. Recently, it is proved that the first two patterns are dominated both in \citet{2012ApJ...756..112L} and \citet{2018arXiv181007313Y}. However, it is not very \edit1{\replaced{clearly}{clear}} for the physical origin of these two patterns. On the other hand, the correlation analysis for \edit1{\replaced{the parameters evolution}{the evolution of the parameters}} in a single burst is lacking \edit1{\deleted{of}}, except for GRB 131231A in \citet{2019arXiv190104925L} which is a single-pulse burst.

Additionally, since the statistics given by Compton Gamma Ray Observatory told ones that, maybe, it's originated from cosmology for GRBs \citep{1992Natur.355..143M}. And there are \edit1{\replaced{much}{many}} more astrophysical scientists or astronomical workers take it as a career in their life to explore the universe probed by GRBs. So, more and more Gamma-Ray Monitors are born in these years. Especially, the launch of \edit1{\added{the}} Swift Gamma-Ray Burst Mission \edit1{\added{($Swift$)}} with three instruments \citep{2004ApJ...611.1005G} gave birth to a new era of GRBs. The broad-band afterglow light curves were recorded by the X-ray telescope (XRT) and the UV/optical telescope onboard $Swift$. It makes one \edit1{\replaced{sieze}{seize}} a chance to summarize the \edit1{\replaced{stantard}{standard}} X-ray afterglow light curve such as \citet{2006ApJ...642..354Z}. And the launch of \edit1{\added{the}} Fermi Gamma-ray Space Telescope \edit1{\added{($Fermi$)}} in 2008 \citep{2009ApJ...697.1071A} make it possible to detect GRBs in broad energy range both in \edit1{\added{the}} prompt emission and afterglow emission.
In fact, there are some GeV photons were detected in some GRBs in \edit1{\added{the}} prompt emission or afterglow emission such as GRB 130427A with GeV photons both in \edit1{\added{the}} prompt emission and afterglow \citep{2014Sci...343...42A}, GRB 940217, a burst that the highest energy photon was detected with \edit1{\added{the}} energy of 18 GeV  in \edit1{\added{the}} afterglow \citep{1994Natur.372..652H}. Some focused on \edit1{\added{the}} highest energy photons in \edit1{\added{the}} prompt emission like \citet{2005ApJ...622L..25T}. And others focused on those highest energy photons in \edit1{\added{the}} afterglow with \edit1{\added{the}} interpretation of synchrotron radiation \citep{2009MNRAS.400L..75K,2009ApJ...706L..33G,2010MNRAS.403..926G,2011ApJ...733...22H,2013ApJ...763...71A,2014ApJ...781...74L} very early, but, another named inverse-Compton (IC) emission as the leading mechanisms for GeV photons was accepted such as \citet{2014ApJ...787L...6L} and \citet{2014Sci...343...42A} since the hypothesis of IC in GeV emission was given by \cite{2009MNRAS.396.1163Z}. To our surprise, in \edit1{\added{the}} afterglow with GeV emission, there are two types of inverse-Compton emission with GeV emission during X-ray flares, synchrotron self-Compton (SSC) radiation \edit1{\replaced{arised}{arose}} from the interaction between photons and electrons that \edit1{\replaced{creat this photons}{create these photons}} \citep{1994MNRAS.269L..41M}, and another one is external inverse Compton (EIC) emission due to the interaction between photons and hot electrons in external shock \citep{2008MNRAS.384.1483F}.

GRB 180720B, the gamma-ray burst observed by $Fermi$ (co-detected by \edit1{\added{the}} \edit1{\replaced{GBM}{Gamma-ray Burst Monitor (GBM)}} and \edit1{\replaced{LAT}{Large Area Telescope (LAT)}}) and $Swift$ (co-detected by \edit1{\added{the}} \edit1{\replaced{BAT}{Burst Alert Telescope (BAT)}} and \edit1{\replaced{XRT}{X-Ray Telescope (XRT)}}) recently, is a long, extremely bright and peculiar gamma-ray burst. \edit1{\deleted{And the GBM light curve is a multipulse structure consists of an initial continuedly multipeaked emission episode lasting for a dozen seconds ($\sim$ 8-20 s), a sharp pulse with lower amplitude at about 30 seconds after trigger, and another sharp pulse with lowest amplitude at about 50 s, which fades to a weaker level after $\sim$ 200 s.}}{\explain{Same description as below}}\edit1{\replaced{Since the publishment of data that observed by telescopes such as $Swift$ and $Fermi$, it's not impossible that we can perform the detailed time-resolved spectra analysis and in afterglow even at gamma-ray band.}{We can perform the detailed analysis both for the prompt and afterglow phase with the public data.}} Especially, it's very interesting that the $\alpha$ and $E_{p}$ exhibit the `flux-tracking' \edit1{\replaced{behaviour}{behavior}} and \edit1{\replaced{it's very surprising that}{surprisingly,}} the highest photons with energy of GeV were found during \edit1{\added{the}} X-ray flare in \edit1{\added{the}} early afterglow, which is different from GRB 100728A in mechanism (see \citet{2011ApJ...734L..27A}). 

In this work, \edit1{\replaced{after performing the detailed time-resolved spectral analysis of the multipeaked pulse in the bright gamma-ray burst named GRB 180720B in prompt phase}{after perfoming the temporal characteristics analysis and detailed time-resolved spectral analysis of the multipeaked pulse in the prompt phase in the bright gamma-ray burst named GRB 180720B}}, we give the $E_{p}$ and $\alpha$ evolution patterns. The parameters correlations also will be presented in our analysis. We also \edit1{\replaced{analyse}{analyze}} the \edit1{\replaced{lightcurves}{light curves}} and \edit1{\replaced{spectra}{spectrum}} during \edit1{\added{the}} X-ray flare in \edit1{\added{the}} afterglow of this burst by using $Fermi$ data \edit1{\deleted{of GRB 180720B}} include GBM and LAT energy band. At the same time, the public $Swift$/XRT data was also used to \edit1{\replaced{analyse}{analyze}} the \edit1{\replaced{spectra}{spectral}} components of afterglow emission because there would be \edit1{\replaced{a}{an}} interesting result in the radiation mechanism for gamma-ray afterglow (100 MeV to GeV) if we take it \edit1{\added{to}} account for the  mechanism in \edit1{\added{the}} afterglow. And then we discuss its physical interpretation (synchrotron origin or photosphere model) both for the $E_{p}$ and $\alpha$ evolution pattern with the consideration that the low-energy \edit1{\replaced{powerlaw}{power law}} index $\alpha$ does not exceed the synchrotron limits given by \citet{1998ApJ...497L..17S} and we give the reasonable interpretation of \edit1{\added{the}} mechanism for very high energy gamma-ray afterglow.

\section{GRB 180720B} \label{sec:GRB 180720B}
The Fermi Gamma-Ray Burst Monitor (GBM) triggered on GRB 180720B (trigger 553789304/180720598) at 14:21:39.65 UT ($T_{0}$) on 20 July 2018 and its duration ($T_{90}$) is 49 s with the observation from 50 to 300 keV \citep{2018GCN.22981....1R}. Then, Fermi-LAT detected high energy emission from the burst at 14:21:44.55 according to \citet{2018GCN.22980....1B}. In addition to the $Fermi$, $Swift$/BAT triggered the burst at 14:21:44 UT on 20 July 2018 (trigger=848890) \citep{2018GCN.22973....1S} with a duration ($T_{90}$) of 108 s (\url{http://gcn.gsfc.nasa.gov/notices_s/848890/BA/}). Similarly, $Swift$/XRT detected the signal from it (86.5 s after the BAT trigger), and there is a bright flare in \edit1{\added{the}} X-ray band with a duration of more than one hundred seconds \citep{2018GCN.22973....1S}. \edit1{\deleted{So, it's a gamma-ray burst co-detected by $Fermi$ (GBM and LAT) and $Swift$/BAT.}} And one deduced that the redshift of the burst is z=0.654 since several absorption features are detected at z=0.654 in \citet{2018GCN.22996....1V}.

In the $Fermi$ data, there are some interesting signatures. For the GBM data, the fluence is $2.985\pm0.001\times10^{-4} erg/cm^{2}$ in the 10 keV to 1000 keV energy range over the $T_{90}$ interval  \citep{2018GCN.22981....1R}. The time-averaged spectrum (from trigger to 55 s after trigger) is best fit by the Band function \citep{1993ApJ...413..281B} with $E_{peak}=631\pm10$ keV, $\alpha=-1.11\pm0.01$, $\beta=-2.30\pm0.03$ \citep{2018GCN.22981....1R} in GBM energy range. It's very bright in LAT energy range and the \edit1{\replaced{most highest energy photon}{highest-energy photon}} is detected at 137 seconds after the GBM trigger with \edit1{\added{the}} energy of 5 GeV according to  \citet{2018GCN.22980....1B}, but, in fact, we found that this photon was observed by LAT at 142 s after \edit1{\added{the}} GBM trigger through making likelihood analysis (see Section \ref{subsubsec:3.2.1}). Such a photon with \edit1{\added{the}} highest energy in \edit1{\added{the}} afterglow is very monstrous like GRB 130427A \citep{2014Sci...343...42A}.

In the $Swift$ data, the observation of BAT shows a multi-peaked structure from trigger to about 150 s after \edit1{\added{the}} trigger \citep{2018GCN.22973....1S}. The fluence is $8.6\pm0.1\times10^{-5} erg/cm^{2}$ in the 15 keV to 150 keV energy range \citep{2018GCN.22998....1B} according to BAT data. Then, we can find that there is a very bright flare as said above with a duration of more than one hundred seconds \citep{2018GCN.22973....1S} in \edit1{\added{the}} X-ray band. If we are careful enough, the phenomenon would be  found easily that there are some GeV photons while the X-ray flare was detected (see \edit1{\deleted{Figure \ref{fig:lightcurve_all} and Figure \ref{fig:lightcurve_xrt} in}}Section \ref{subsubsec:3.2.1}).

\section{Data analysis of GRB 180720B and the results} \label{sec:Data analysis}

\subsection{Analysis in Prompt Phase} \label{subsec:3.1}

\subsubsection{ \color{red} Temporal Characteristics and Time-resolved Spectral Analysis}
 \label{subsubsec:3.1.1}

\begin{figure}[ht!]
\plotone{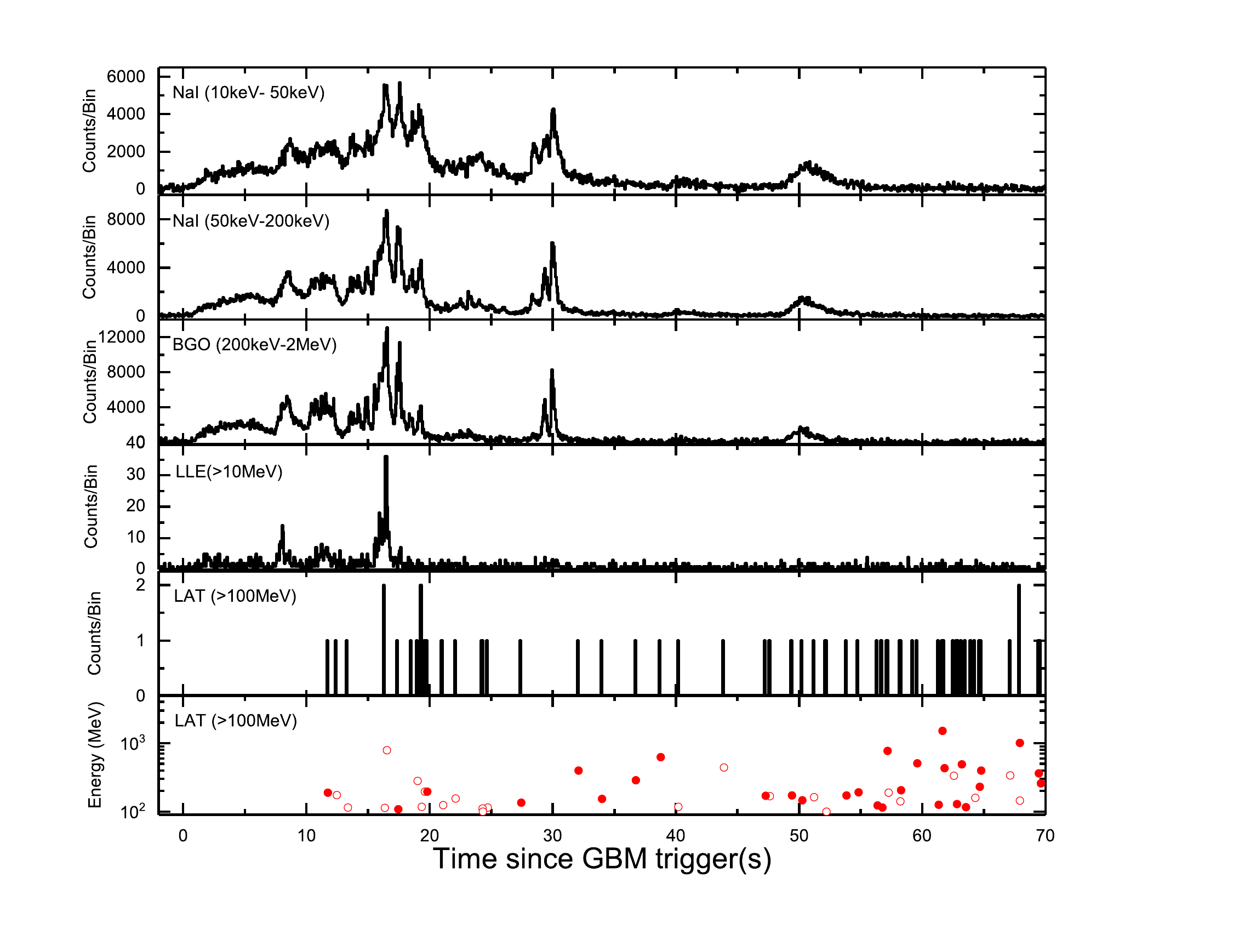}
\caption{\edit1{\added{Light curve for the Fermi-GBM and LAT detectors during the prompt emission in 0.064 s bins, divided into 5 energy ranges. We used the CTIME type of GBM data included NaI detectors 6, 7 and BGO detector 1. On the bottom, the open circles represent the individual LAT `transient' class photons and their energies; the red solid circles indicate those photons with a $\geq 0.9$ probability of being associated with GRB 180720B.}}}\explain{This picture was added in the first revision.}
 \label{fig:lightcurve}
\end{figure}

\begin{figure}
\plotone{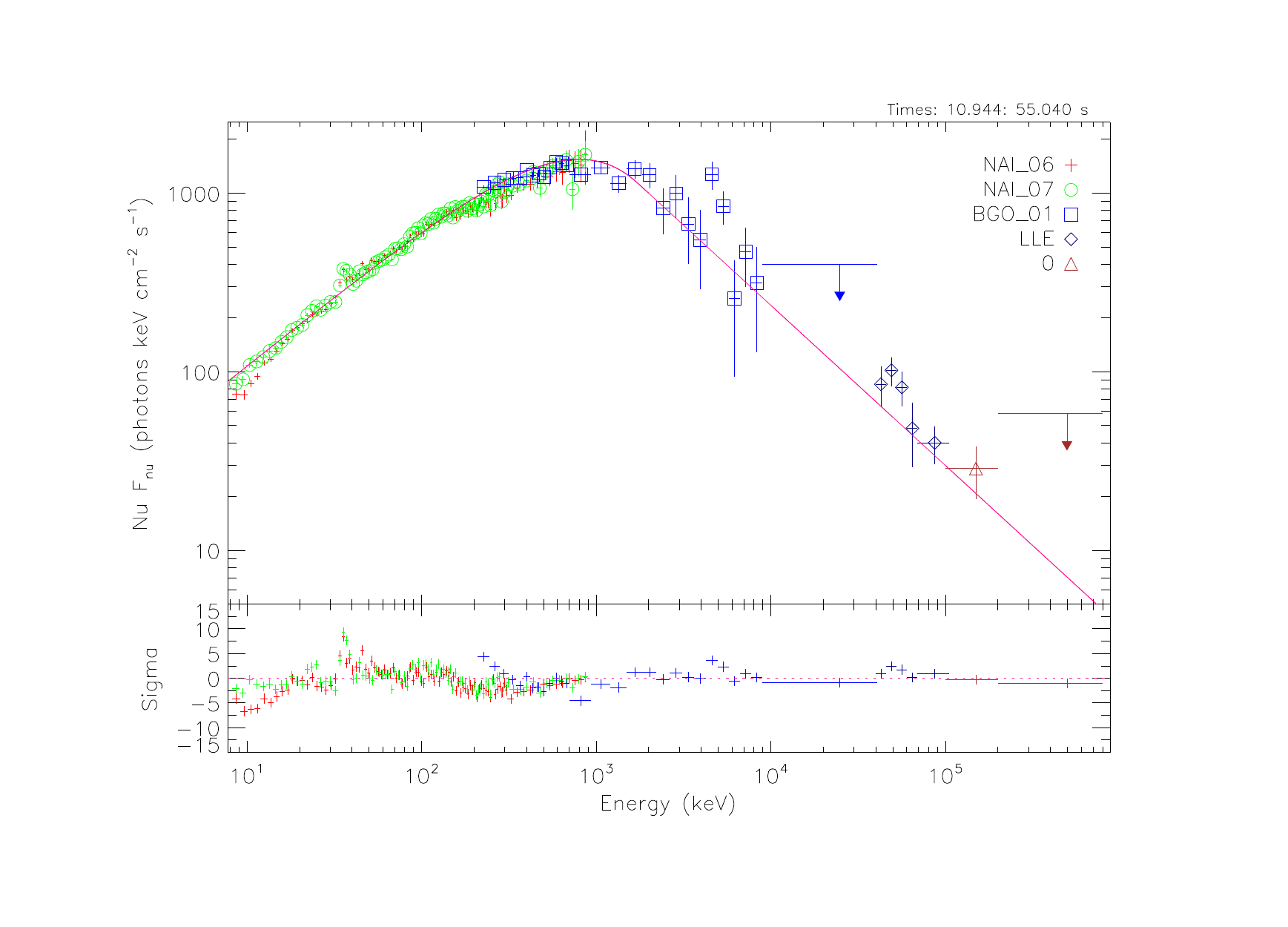}
\caption{\edit1{\added{The joint GBM and LAT (include LAT-LLE) spectrum from 11 s to 55 s in the prompt phase. The Band function was adopted to make the best-fitting, with $\alpha\sim -1.21$, $\beta\sim-2.89$, $E_{p}\sim800$ keV. The deep-pink solid line shows the best-fitting result. The energy range from 8 keV to 1 GeV.}\explain{This picture was added in the first revision.}}}\label{fig:spectrum1} 
\end{figure}

The $Fermi$ data of GRB 180720B that we utilized are available at the Fermi Science Support Center (FSSC).\footnote{\url{https://fermi.gsfc.nasa.gov/ssc/data/access/}} \edit1{\deleted{But we gave up the use of LAT data due to the fewer photons.}} We extract GBM \edit1{\replaced{lightcurve}{light curve}} from the TTE (Time-Tagged Events) data by using a Python source package named gtBurst.\footnote{\url{https://github.com/giacomov/gtburst}} \edit1{\added{For the LAT data, we also used the gtBurst code to make unbinned maximum likelihood analysis. According to \citet{2019ApJ...878...52A}, we selected the Pass 8 P8R2$\_$TRANSIENT020E$\_$V6 event class and the corresponding response function for the time window starting at the trigger time $T_{0}$ and ending at 100 s after the GBM trigger (bottom panel in Figure \ref{fig:lightcurve}). Another time window starting at the trigger time $T_{0}$ and ending at 10000 s after the GBM trigger, we selected the Pass 8 P8R2$\_$TRANSIENT010E$\_$V6 event class and the corresponding response function (Table \ref{tab:likelihood}). We consider an ROI centered on with a radius of $12^{\circ}$ from 100 MeV to 5 GeV. Those photons with the zenith angle smaller than $100^{\circ}$ were considered to reduce the contamination of the gamma photons from the earth limb. Then we run the tool $gtsrcprob$ to compute the probability of being associated with GRB 180720B for each photon. All of the photons which have a probability larger than $90\%$ have been presented in the bottom panel in Figure \ref{fig:lightcurve}.}} To complete this analysis, we also take RMFIT as the tool of making \edit1{\added{the}} spectral analysis. We perform the \edit1{\replaced{spetral}{spectral}} analysis by using the data of two NaI detectors ($n_{6}$, $n_{7}$) and one BGO detector ($b_{1}$) on $Fermi$/GBM. The energy range for each spectrum \edit1{\replaced{coverd}{covered}} from 10 keV to 40 MeV. And the background photon counts were estimated by fitting the \edit1{\replaced{lightcurve}{light curve}} before and after the burst with a one-order background polynomial model. We selected the interval from 0 s to 55 s after \edit1{\added{the}} GBM trigger as the source. The signal-to-noise ratio (S/N) with \edit1{\added{a}} value of 30 was used in all of these slices. \edit1{\added{We gave up the use of LAT data because of its lower impact for peak energy $E_{p}$ and low energy spectral index $\alpha$ besides the fewer photons in the detailed time-resolved spectral analysis, but we make the joint GBM and LAT (include LAT-LLE) spectrum from 11 s to 55 s in the prompt phase (see Figure \ref{fig:spectrum1}).}} And they all can be fitted well by Band \citep{1993ApJ...413..281B}. The goodness-of-fit in our analysis by reduced $\chi^{2}$. The \edit1{\replaced{best fitting}{best-fitting}} results for all of these slices are presented in Table \ref{tab:resolved_results}.

\edit1{\added{For GRB 180720B, the temporal profile of the emission  varies with energy from 10 keV to 5 GeV (Figure \ref{fig:lightcurve}). The GBM light curve is a multipulse structure consists of an initial continuedly multipeaked emission episode lasting for a dozen seconds ($\sim$8-20 s), a sharp pulse with a lower amplitude at about 30 s after the trigger, and another sharp pulse with the lowest amplitude at about 50 s in the prompt phase (Figure \ref{fig:lightcurve}). The LLE light curve is sharper than the other detectors-detected emission although it also exhibits a multipeaked structure. They have counterparts in the GBM energy range for these peaks. However, the LAT-detected emission is the weakest which has fewer photons with a $\geq 0.9$ probability of being associated with this burst.}}

\begin{deluxetable}{ccccc}
\tablewidth{0pt}
\renewcommand\tabcolsep{20.0pt}
\tabletypesize{\footnotesize}
\tablecaption{\textbf{\color{red} Result of Unbinned Maximum Likelihood Analysis in Each Time Interval for GRB 180720B}}
\tablehead{
\colhead{Time Interval(s)} & \colhead{Energy Flux\tablenotemark{a}} & \colhead{Photon Flux\tablenotemark{b}} & \colhead{Photon Index} & \colhead{TS value} 
}				
\startdata
11--60&(5.99$\pm$1.53)$\times$10$^{-8}$&(2.2$\pm$0.56)$\times$10$^{-4}$&-3.41$\pm$0.48&180\\
60--90&(1.35$\pm$0.43)$\times$10$^{-7}$&(2.32$\pm$0.67)$\times$10$^{-4}$&-2.10$\pm$0.29&172\\
90--105&(1.80$\pm$0.68)$\times$10$^{-7}$&(3.43$\pm$1.20)$\times$10$^{-4}$&-2.20$\pm$0.37&120\\
105--145&(1.17$\pm$0.40)$\times$10$^{-7}$&(1.26$\pm$0.41)$\times$10$^{-4}$&-1.67$\pm$0.28&141\\
145--250&(4.19$\pm$1.13)$\times$10$^{-8}$&(7.29$\pm$1.76)$\times$10$^{-5}$&-2.11$\pm$0.24&185\\
250--550&(6.79$\pm$2.58)$\times$10$^{-9}$&(1.17$\pm$0.39)$\times$10$^{-5}$&-2.09$\pm$0.34&68\\
550--1000&6.95$\times$10$^{-9}$&1.09$\times$10$^{-5}$&--&11\\
4500--6900&1.56$\times$10$^{-9}$&2.44$\times$10$^{-6}$&--&10\\
\enddata
\tablenotetext{a}{In the unit of $ergs{\cdot}cm^{-2}s^{-1}$; values without uncertainty are upper limits.}
\tablenotetext{b}{In the unit of $photons{\cdot}cm^{-2}s^{-1}$; values without uncertainty are upper limits.}\label{tab:likelihood}

\explain{We added this table in the first revision.}
\end{deluxetable}

\edit1{\added{In the unbinned maximum likelihood analysis (Table \ref{tab:likelihood}), the photons were divided into 8 time intervals. We added the ``Galactic diffuse" and ``isotropic diffuse" components. A single power-law spectrum with its ``normalization" and ``spectral index" being allowed to vary was assumed for this burst. The TS value in each time interval is larger than 9. The fact that the TS values are very large indicates that most of the photons are associated with the burst in our analysis.}}

\edit1{\added{As we can see from Figure \ref{fig:lightcurve}, there are no photons with energies greater than 1 GeV from 11 s to 55 s in the prompt emission for GRB 180720B. With the photon statistics permission for the LAT data, we make the joint GBM and LAT (include LAT-LLE) time-averaged $\nu F_{\nu}$ spectrum in this time interval (Figure \ref{fig:spectrum1}). We selected the NaI detectors with the energy range from $\sim 8$ keV to 900 keV, the BGO detector with the energy range from 200 keV to 40 MeV, the LLE detector from 40 MeV to 100 MeV, the LAT detector from 100 MeV to 1 GeV, which indicate the energy range covered from 8 keV to 1 GeV completely. The Band function with $\alpha \sim -1.21$, $\beta \sim -2.89$, $E_{p} \sim 800$ keV described the spectrum, which means that they share a common origin for low energy emission and high energy emission in the prompt phase.}}

\begin{deluxetable}{ccccccc}
\tablecaption{Results of the Time-resolved Spectral Fits for GRB 180720B \label{tab:resolved_results}}
\tablehead{
\colhead{$t_{1} \sim t_{2}$ \tablenotemark{a}} 
& \colhead{$\alpha$}
& \colhead{$\beta$}
& \colhead{$E_{p}$}
& \colhead{flux}
& \colhead{$\chi^{2}/dof$}
& \colhead{Red.$\chi^{2}$}\\
\colhead{s}& \colhead{}& \colhead{}& \colhead{keV} & 
\colhead{$\times10^{-6}erg/cm^{2}/s$} & \colhead{} & \colhead{}
}
\colnumbers
\startdata
3.552$\sim$5.467&-0.95$\pm$0.02&-2.31$\pm$0.09&1048$\pm$85.5&8.5$\pm$0.11&373.70/355&1.05\\
5.467$\sim$7.524&-0.97$\pm$0.02&-2.20$\pm$0.08&708.5$\pm$56.9&7.1$\pm$0.12&398.45/355&1.12\\
7.524$\sim$8.591&-0.90$\pm$0.02&-2.12$\pm$0.06&965.7$\pm$72.0&15$\pm$0.19&342.18/355&0.96\\
8.591$\sim$10.223&-1.01$\pm$0.02&-2.20$\pm$0.07&616.8$\pm$40.4&11$\pm$0.15&415.64/355&1.17\\
10.223$\sim$11.667&-0.95$\pm$0.01&-2.36$\pm$0.07&953.4$\pm$53.2&16$\pm$0.16&393.96/355&1.11\\
11.667$\sim$13.306&-1.00$\pm$0.02&-2.17$\pm$0.06&569.3$\pm$37.5&10$\pm$0.15&385.08/355&1.08\\
13.306$\sim$14.713&-0.94$\pm$0.02&-2.19$\pm$0.06&432.7$\pm$25.4&11$\pm$0.15&469.61/355&1.32\\
14.713$\sim$15.029&-0.94$\pm$0.03&-2.58$\pm$0.22&719.6$\pm$68.5&17$\pm$0.38&320.21/355&0.90\\
15.029$\sim$15.415&-0.89$\pm$0.05&-2.03$\pm$0.08&310.2$\pm$35.1&11$\pm$0.24&371.88/355&1.05\\
15.415$\sim$16.024&-0.90$\pm$0.02&-2.28$\pm$0.07&782.7$\pm$50.0&23$\pm$0.31&445.08/355&1.25\\
16.024$\sim$16.597&-0.84$\pm$0.02&-2.12$\pm$0.04&838.3$\pm$41.2&41$\pm$0.42&438.27/355&1.23\\
16.597$\sim$17.257&-0.96$\pm$0.02&-2.08$\pm$0.05&526.8$\pm$35.2&21$\pm$0.29&387.41/355&1.09\\
17.257$\sim$17.649&-0.86$\pm$0.02&-2.53$\pm$0.10&685.2$\pm$35.8&35$\pm$0.47&402.40/355&1.13\\
17.649$\sim$18.171&-1.04$\pm$0.03&-2.17$\pm$0.08&420.7$\pm$35.2&16$\pm$0.27&360.66/355&1.02\\
18.171$\sim$18.979&-1.05$\pm$0.03&-2.31$\pm$0.10&294.2$\pm$20.5&11$\pm$0.17&388.69/355&1.09\\
18.979$\sim$19.358&-1.12$\pm$0.03&-2.61$\pm$0.25&525.6$\pm$47.4&16$\pm$0.35&338.27/355&0.95\\
19.358$\sim$21.215&-1.25$\pm$0.05&-1.98$\pm$0.05&179.6$\pm$23.5&4.3$\pm$0.073&388.28/355&1.09\\
21.215$\sim$27.937&-1.35$\pm$0.02&-2.12$\pm$0.08&232.3$\pm$21.2&2.8$\pm$0.037&471.65/355&1.33\\
27.937$\sim$29.473&-1.19$\pm$0.02&-2.42$\pm$0.19&459.1$\pm$40.9&6.5$\pm$0.13&343.86/355&0.97\\
29.473$\sim$29.874&-1.00$\pm$0.04&-2.28$\pm$0.15&324.1$\pm$34.5&9.8$\pm$0.24&354.26/355&0.99\\
29.874$\sim$30.055&-0.83$\pm$0.03&-3.11$\pm$0.36&883.2$\pm$64.8&34$\pm$0.63&357.71/355&1.01\\
30.055$\sim$31.524&-1.22$\pm$0.02&-2.89$\pm$0.52&545.7$\pm$47.9&6.9$\pm$0.13&326.67/355&0.92\\
31.524$\sim$48.247&-1.33$\pm$0.06&-1.91$\pm$0.06&161.0$\pm$31.3&0.85$\pm$0.019&393.70/355&1.11\\
48.247$\sim$55.040&-1.16$\pm$0.02&-2.29$\pm$0.15&364.3$\pm$30.8&2.5$\pm$0.042&382.68/355&1.08\\
\enddata
\tablenotetext{a}{Time intervals.}
\end{deluxetable}

\subsubsection{The Peculiar Characteristics of the Spectral Evolution in GRB 180720B: `flux-tracking' patterns for $\alpha$ and $E_{p}$} \label{subsubsec:3.1.2}
As we all know, it may suffer from the influence of the complex central engine for multipulse gamma-ray burst so that many of the properties are harder to \edit1{\replaced{analyse}{analyze}} than the single-pulse. But it's different in our case, for GRB 180720B, which is a multipulse structure \edit1{\deleted{consists of an initial continuedly multipeaked emission episode lasting for dozen seconds ($\sim$8-20 s), a sharp pulse with lower amplitude at about 30 s, and another sharp pulse with lowest amplitude at about 50 s in prompt}}\edit1{\added{described as Section \ref{subsubsec:3.1.1}}}. The temporal evolution of $E_{p}$ and $\alpha$ display \edit1{\replaced{a}{the}} significant `tracking' \edit1{\replaced{trend}{trends}}  along with photon counts, especially, the  `flux-tracking' \edit1{\replaced{trend}{trends}} are also shown in Figure \ref{fig:evolution_patterns} although it's the multipulse burst instead of the single-pulse like GRB 131231A in \citet{2019arXiv190104925L}, which is also the `flux-tracking' pattern \edit1{\added{both}} for its $E_{p}$ and $\alpha$ in all time-resolved spectra.

Recently, the discovery that there are only 8 out of 38 bursts display `flux-tracking' trend for $E_{p}$ was reported in \citet{2018arXiv181007313Y}. On the other hand, they also pointed out that the $\alpha$ evolution has no strong general trend. In general, the multipulse bursts would be more complex if the single-pulse are irregular for the evolution of $E_{p}$ and $\alpha$. However, the fact that both $E_{p}$ and $\alpha$ exhibit the `flux-tracking' patterns suggests that the spectral evolution for GRB 180720B is very peculiar, which is a multipulse structure in \edit1{\added{the}} prompt phase. For the single-pulse, the individual parameter relations will show strong monotonous positive correlations in both the rising and decaying wings since both the $E_{p}$ and $\alpha$ exhibit the `flux-tracking' patterns like GRB 131231A in \citet{2019arXiv190104925L}. So, we guess that the signature will emerge that the individual parameter relations may show \edit1{\deleted{a}} strong monotonous positive correlations across the whole pulse in multipulse GRB 180720B with the `flux-tracking' \edit1{\replaced{behaviour}{behaviors}} of $E_{p}$ and $\alpha$. In the following, the detailed parameter correlation analysis will be present.
\begin{figure}
\gridline{\fig{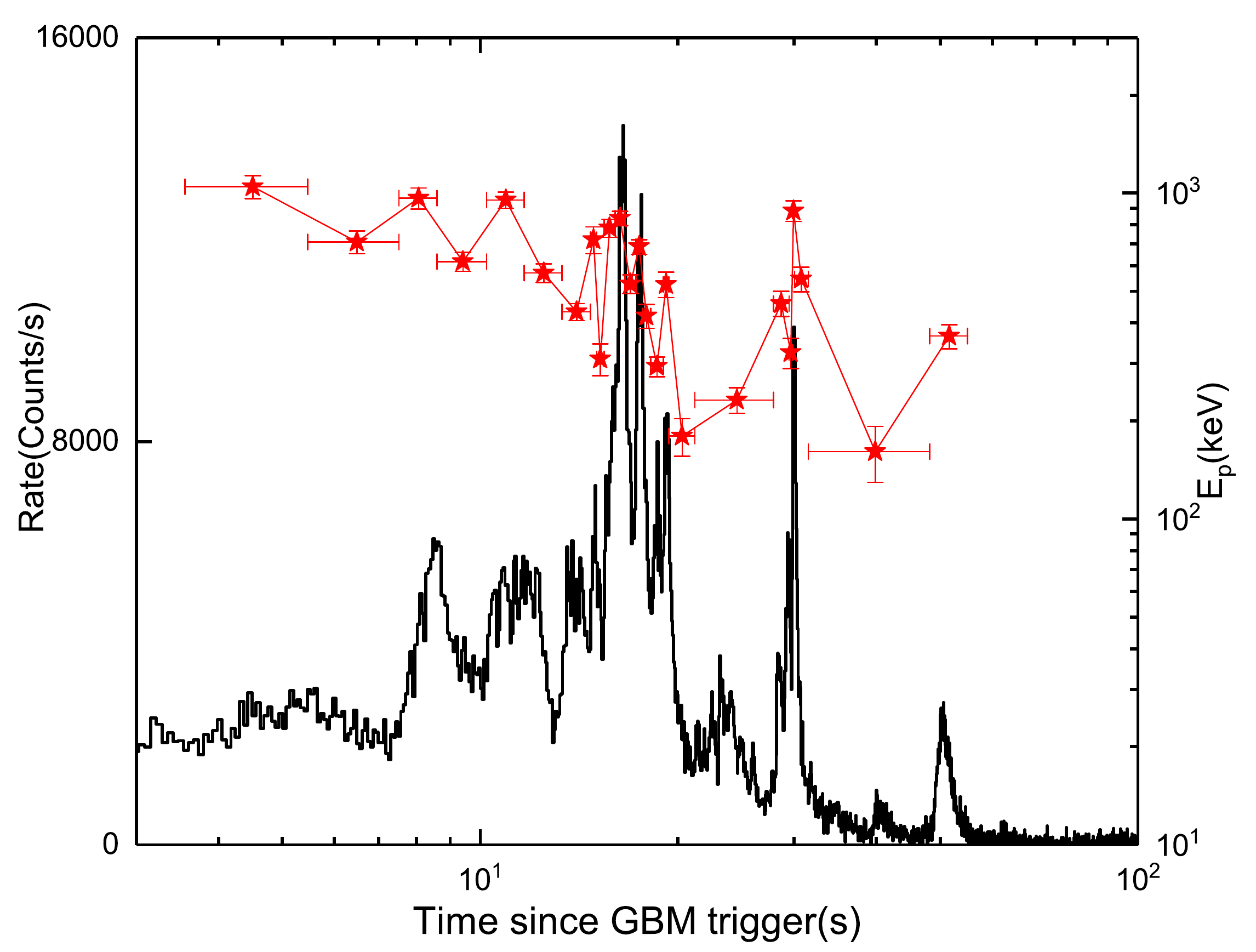}{0.5\textwidth}{}
          \fig{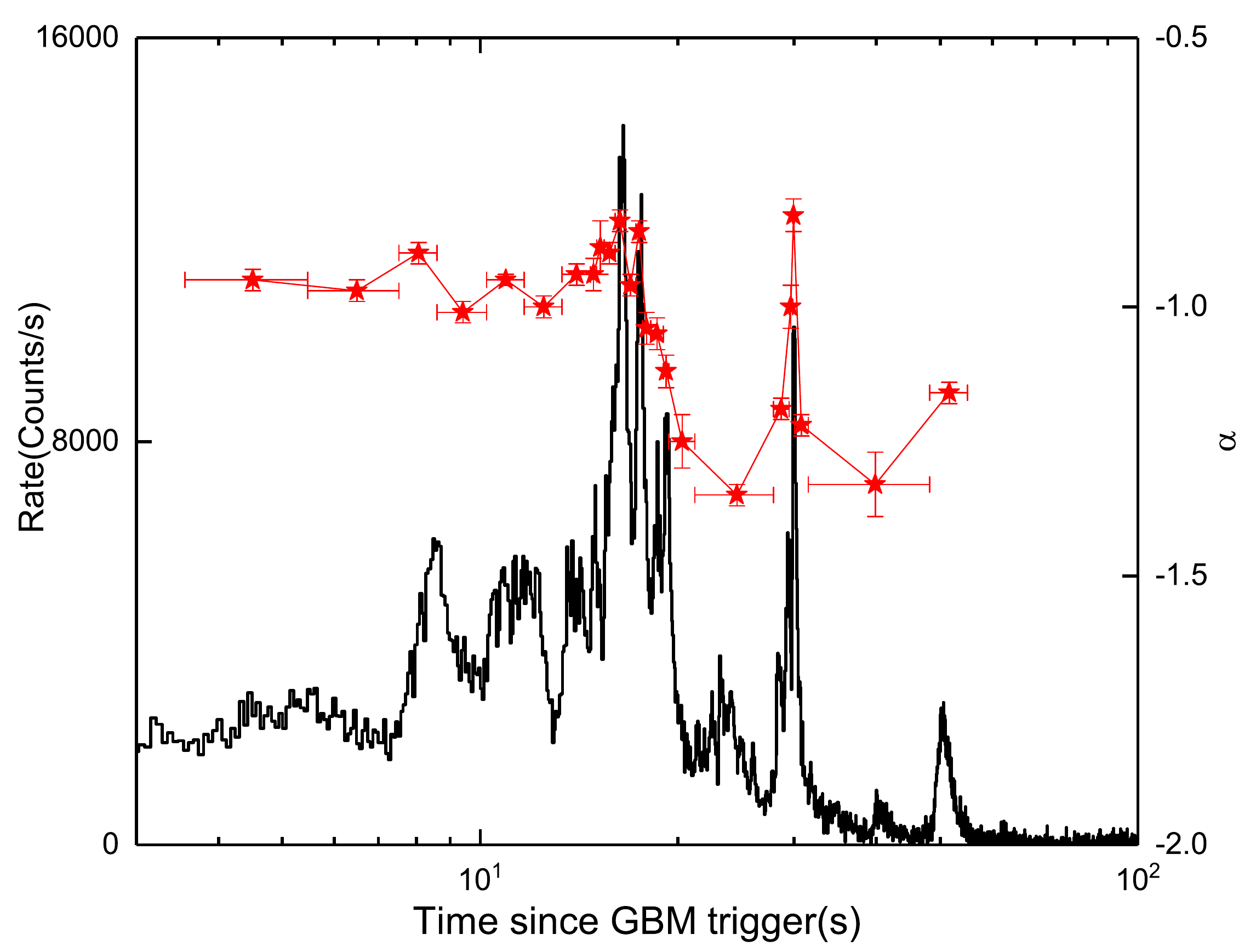}{0.5\textwidth}{}
          }
\gridline{
          \fig{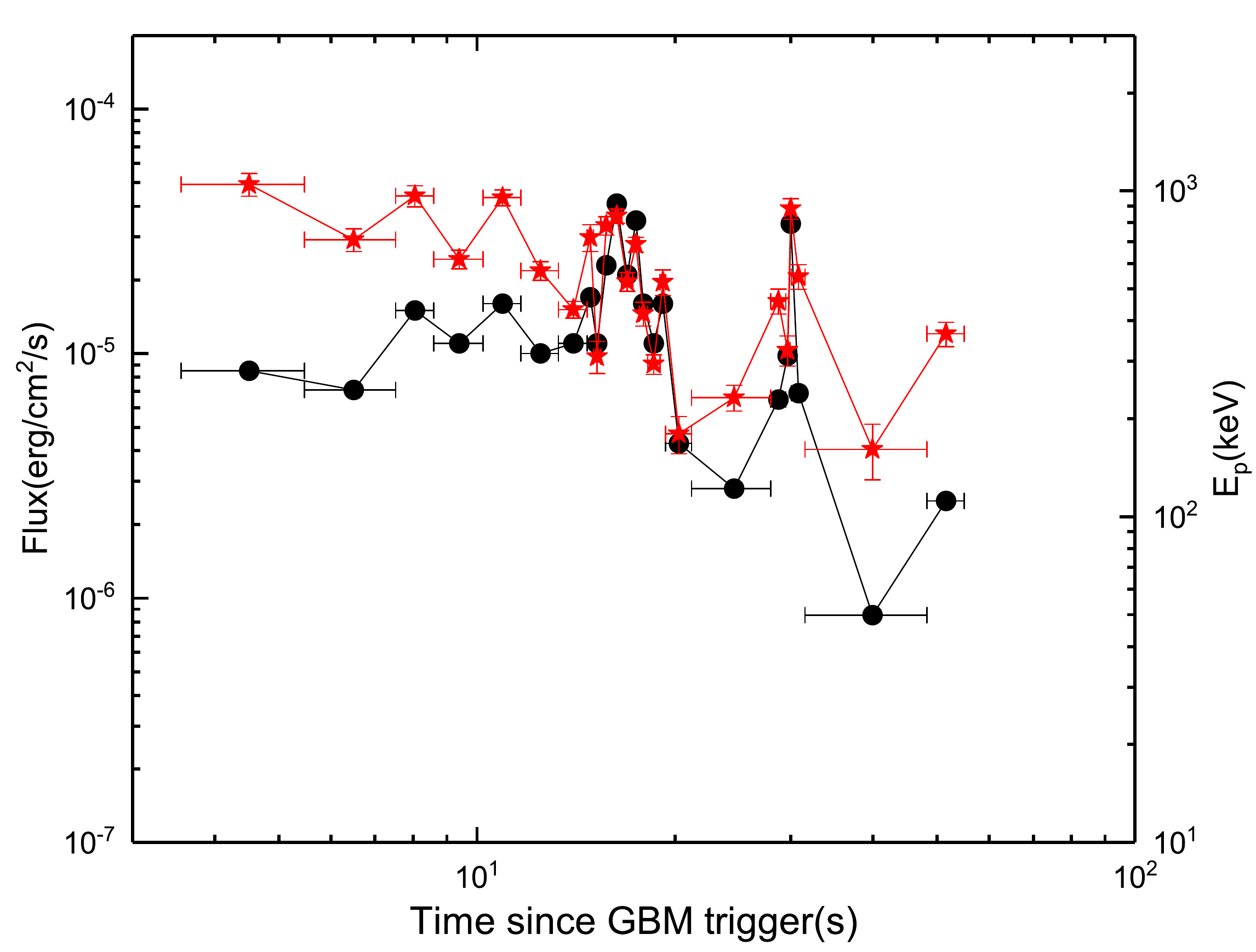}{0.5\textwidth}{}
          \fig{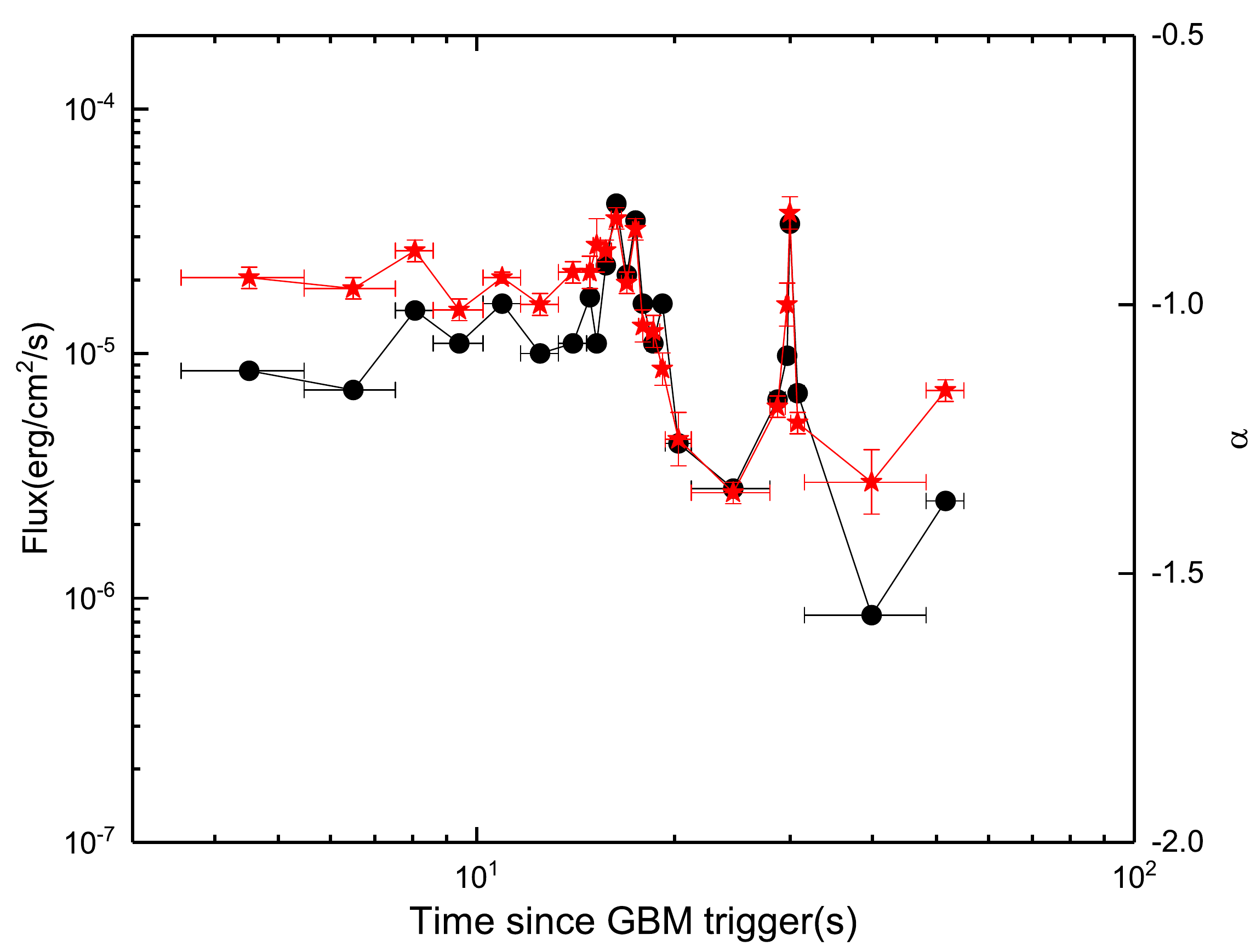}{0.5\textwidth}{}
          }
\caption{The two of top represent \edit1{\added{the}} light \edit1{\replaced{curve}{curves}} of the prompt emission for GRB 180720B (the left-hand y-axis), along with time evolution of the $E_{p}$ (left-panel) and $\alpha$ (right-panel), both are marked with red stars in the right-hand y-axis. The `tracking' signature emerges both for $E_{p}$ and $\alpha$. The two of bottom represent \edit1{\replaced{timeporal}{the temporal}} characteristics of energy flux for GRB 180720B (the left-hand y-axis), along with time evolution of the $E_{p}$ (left-panel) and $\alpha$ (right-panel), both are marked with red stars in the right-hand y-axis. The `flux-tracking' signature emerges both for $E_{p}$ and $\alpha$.}
\end{figure}\label{fig:evolution_patterns}

\subsubsection{Parameter Correlation Analysis} \label{subsubsec:3.1.3}
The parameter correlation analysis is important to reveal the radiation mechanism of GRB in \edit1{\added{the}} prompt. These correlations such as $E_{p}-F$, $\alpha-F$ and $E_{p}-\alpha$ correlations \edit1{\added{are}} shown in Figure \ref{fig:correlation1} and Figure \ref{fig:correlation2}. Besides, another key correlation of $E_{p}-L_{\gamma,iso}$ was also carried in Figure \ref{fig:correlation2}. The time-resolved $E_{p}-F$ in GRB 180720B shows a strong positive correlation through the whole pulse. The best linear fit is 
log$E_{p}$/(keV)=(4.48$\pm$0.46)+(0.35$\pm$0.09)$\times$ log$F$/(erg/cm$^{2}$/s), with number of data points $N=24$, the Pearson's linear correlation coefficient $r=0.62$. The best linear fit is $\alpha$=(0.72$\pm$0.24)+(0.35$\pm$0.05)$\times$ log$F$/(erg/cm$^{2}$/s) ($N=24$, $r=0.84$) for the time-resolved $\alpha-F$ correlation. This $\alpha-F$ relation for GRB 180720B is very similar to the $E_{p}-F$ relation, they show a strong monotonic positive relation. However, it seems that GRB 180720B is a special case in multipulse GRBs, since the fact that the power-law indices for the $\alpha-F$ and $E_{p}-F$ relations are same is quite surprising. 
\edit1{\replaced{In addition}{Besides}}, there are two important relations as shown in Figure \ref{fig:correlation2}. One is the strong monotonic positive relation between $E_{p}$ and $\alpha$, with the best linear fit of log$E_{p}$/(keV)=(3.77$\pm$0.22)+(1.02$\pm$0.22)$\times \alpha$ ($N=24$, $r=0.70$). It's obvious that the value of $\alpha$ does not exceed the synchrotron limits ($-\frac{3}{2}$ to $-\frac{2}{3}$). The known redshift with \edit1{\added{the}} value of $z=0.654$ can make us \edit1{\deleted{to}} calculate the isotropic luminosity for each spectrum. So, \edit1{\deleted{the}} another one is $E_{p}-L_{\gamma,iso}$ relation as shown in \edit1{\added{the}} right panel of Figure \ref{fig:correlation2}. The gray filled circles represent the sample in \citet{2010PASJ...62.1495Y} were carried for comparison. Our best linear fit is \edit1{\replaced{log$E_{p}$/(keV)=(-15.63$\pm$4.94)+(0.35$\pm$0.09)$\times$log$L_{\gamma,iso}$(erg/s) ($N=24$, $r=0.62$)}{log$E_{p}$/(keV)=(-15.91$\pm$4.28)+(0.36$\pm$0.08)$\times$log$L_{\gamma,iso}$(erg/s) ($N=24$, $r=0.68$)}} for GRB 180720B. The Yonetoku's sample gives \edit1{\replaced{log$E_{p}$/(keV)=(-18.23$\pm$1.80)+(0.39$\pm$0.03)$\times$log$L_{\gamma,iso}$(erg/s) ($N=101$, $r=0.75$)}{log$E_{p}$/(keV)=(-24.41$\pm$1.37)+(0.51$\pm$0.03)$\times$log$L_{\gamma,iso}$(erg/s) ($N=101$, $r=0.89$)}}. \edit1{\replaced{Then, it's obvious that the data points exceed the 2$\sigma$ dispersion but in the 3$\sigma$.}{Then, it's obvious that most of the data points do not exceed the 2$\sigma$ dispersion.}}\explain{In the original version, we ignored the applicable conditions of the Yonetoku relation such as the $E_{p}$ is not the observed value but the value in the location of burst and the integrated energy range from 1 keV to 10000 keV for the fluence when we calculate the isotropic luminosity. In this revision, we revised these errors.}

\begin{figure}
\gridline{\fig{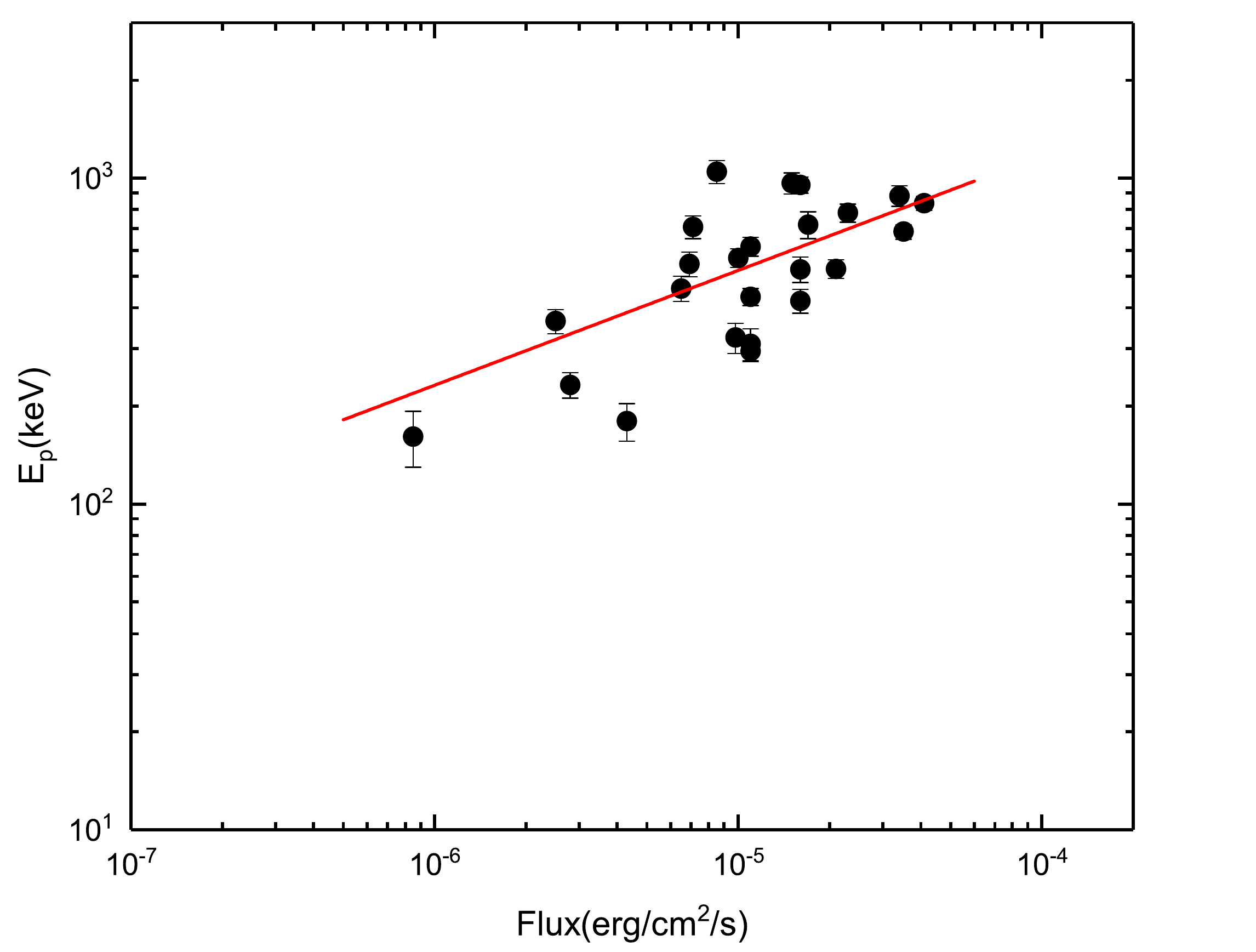}{0.5\textwidth}{}
          \fig{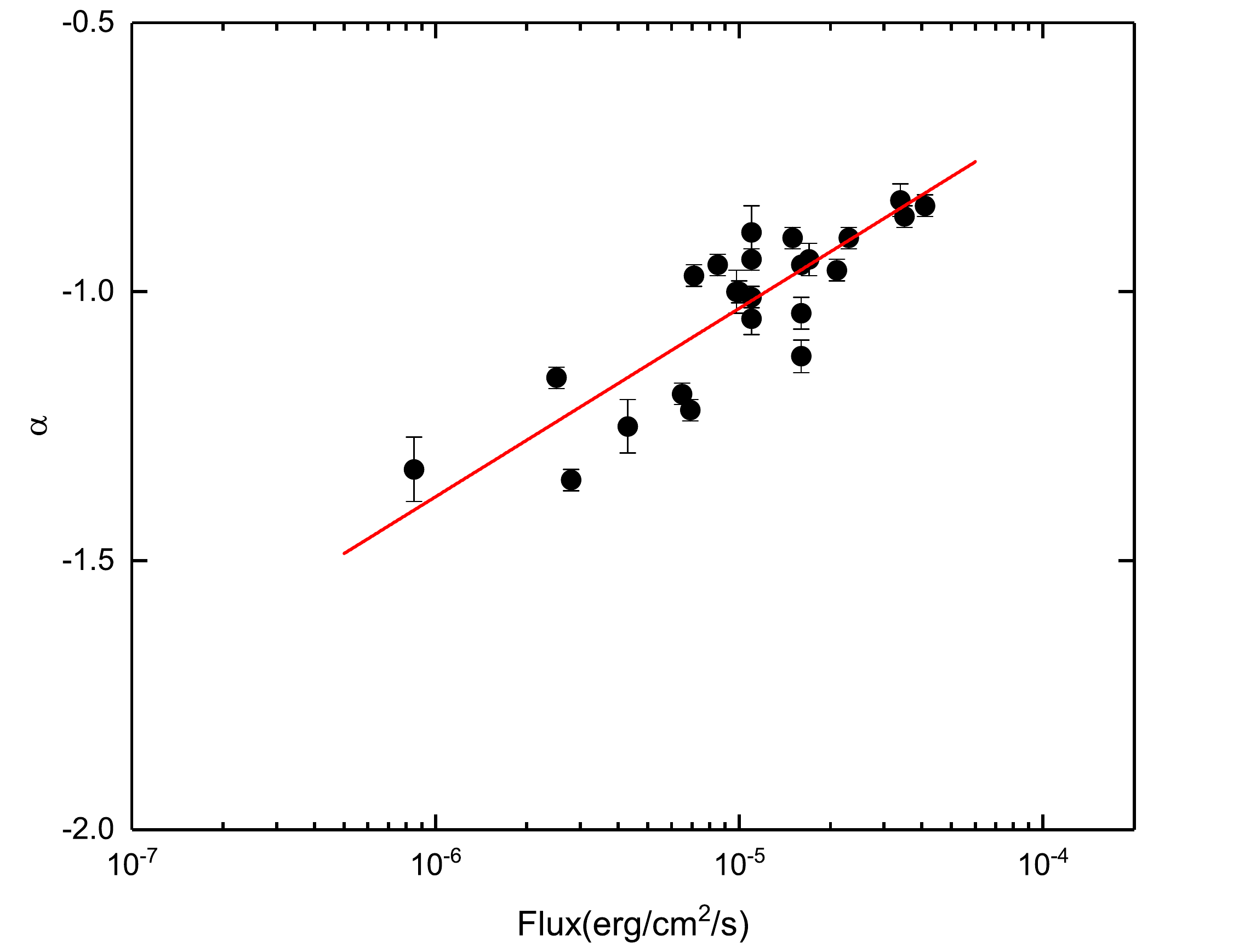}{0.5\textwidth}{}
          }
\caption{$E_{p}$ (left panel) and $\alpha$ (right panel) as a function of flux during the interval from 0 s to 55 s after trigger. The red solid lines represent the best fit for them.}
\end{figure}\label{fig:correlation1}

\begin{figure}
\gridline{\fig{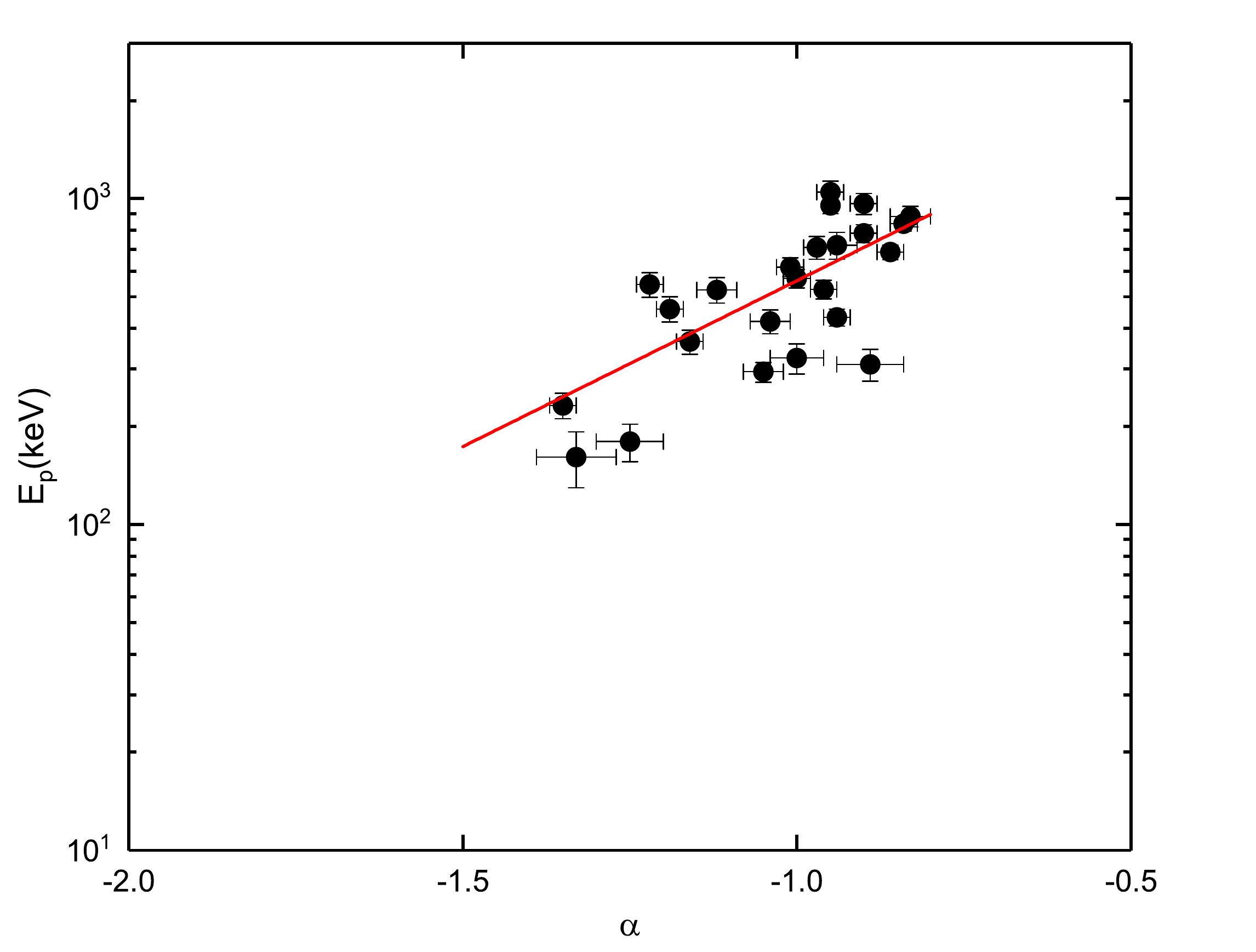}{0.5\textwidth}{}
          \fig{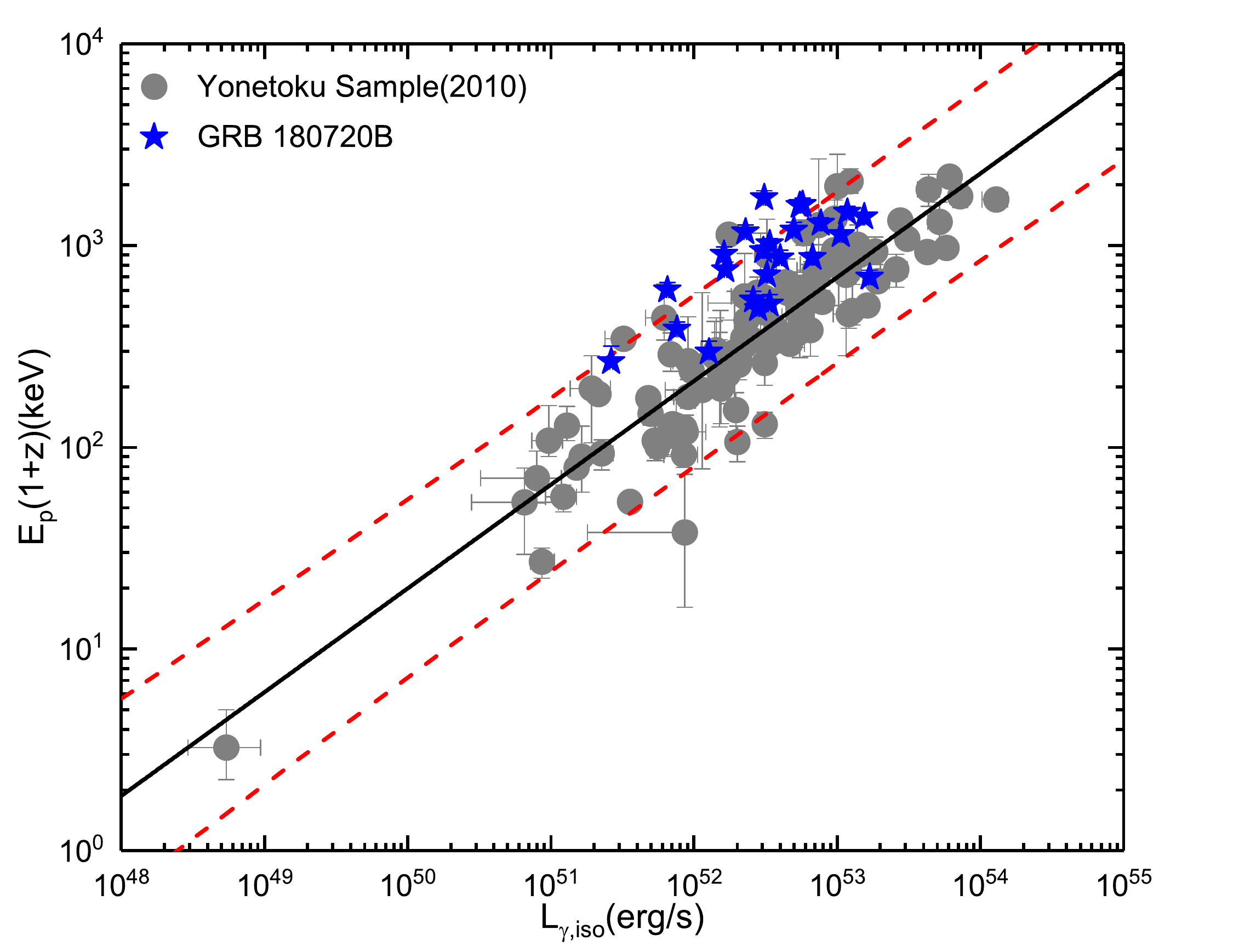}{0.5\textwidth}{}
          }
\caption{The left panel shows the $E_{p}$-$\alpha$ relation. It's obvious that the value of $\alpha$ does not exceed the synchrotron limits ($-\frac{3}{2}$ to $-\frac{2}{3}$). The right panel shows the \edit1{\replaced{comparision}{comparison}} of the time-resolved $E_{p}$-$L_{\gamma,iso}$ correlation for the whole interval of GRB  180720B (blue) with the time-integrated $E_{p}$-$L_{\gamma,iso}$ correlation for the 101 GRBs in \citet{2010PASJ...62.1495Y} (gray filled circles). The black solid line is the best fit to the time-integrated spectra of Yonetoku sample, \edit1{\deleted{while the two black shotdot lines represent its 2$\sigma$ dispersion around it, the two red dash lines represent its 3$\sigma$ dispersion around it.}} \edit1{\added{while the two red dash lines represent its 2$\sigma$ dispersion around it.}}}\explain{In the original version, we ignored the applicable conditions of the Yonetoku relation such as the $E_{p}$ is not the observed value but the value in the location of burst and the integrated energy range from 1 keV to 10000 keV for the fluence when we calculate the isotropic luminosity. In this revision, we revised these errors.}
\end{figure}\label{fig:correlation2}

\subsection{Analysis During High Energy Emission Phase} \label{subsec:3.2} 
As said in Section \ref{subsubsec:3.1.1}, the $Fermi$ data of GRB 180720B that we utilized are available at the Fermi Science Support Center (FSSC). We also extract the LAT light curve and spectrum by using a Python source package named gtBurst. \edit1{\deleted{For LAT data, we selected the events with source class from 100 MeV to 100 GeV within 15$^{\circ}$ around the circle center from FSSC.}}\edit1{\added{We used the same method with Section \ref{subsubsec:3.1.1} include the same parameters setting.}} The $Swift$/XRT light curve and spectra are taken from the $Swift$ Analyzer.\footnote{\url{http://www.swift.ac.uk/burst_analyser/00848890/}} To complete this analysis, we also take RMFIT and Xspec as the tools of making \edit1{\added{the}} spectral analysis.

\subsubsection{Temporal Analysis} \label{subsubsec:3.2.1}

\begin{figure}
\plotone{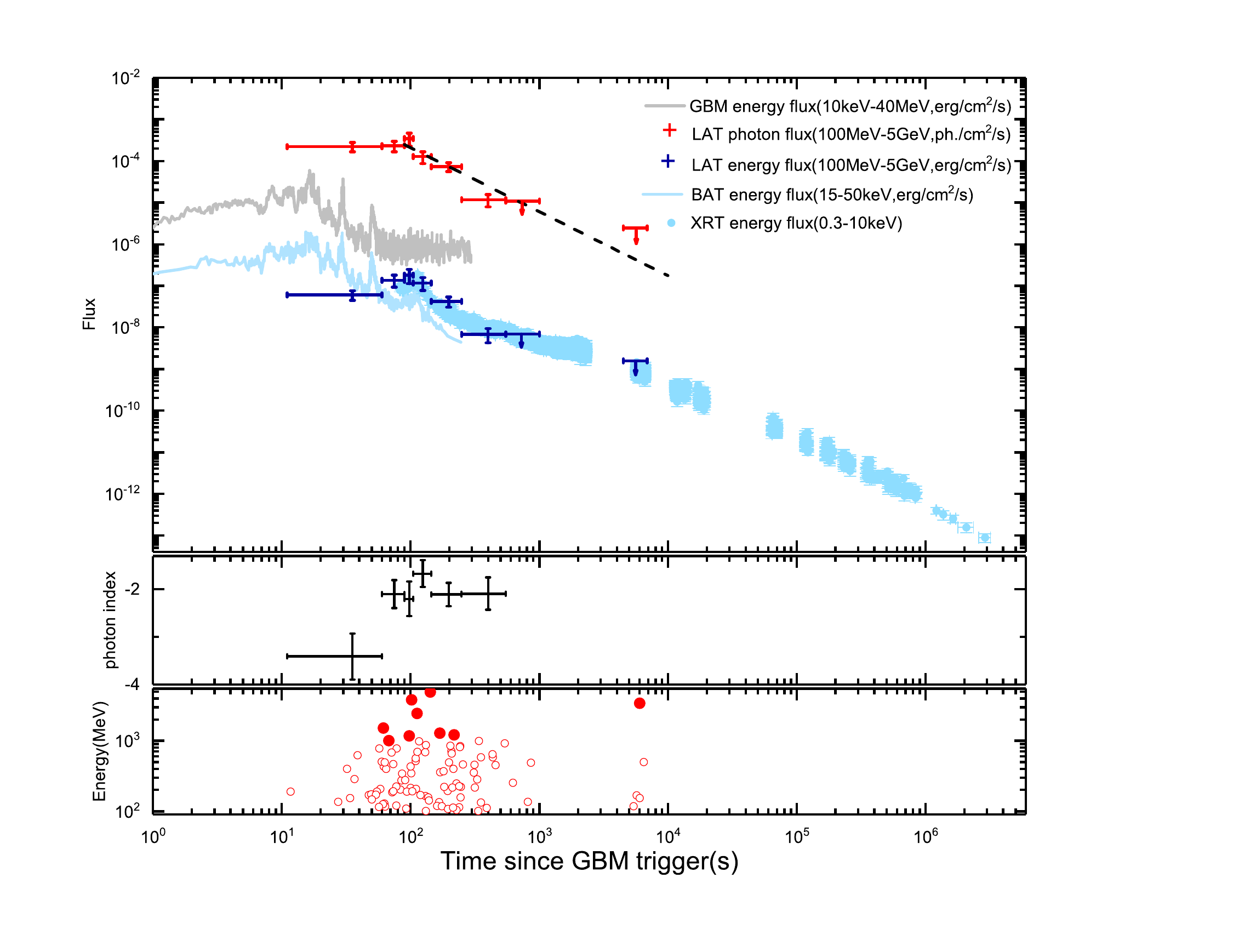}
\caption{Temporal characteristics. The upper panel shows that multi-wavelength light curves from \edit1{\added{the}} prompt to afterglow phase of GRB 180720B from \edit1{\added{the}} X-ray band to gamma-ray energy range. \edit1{\added{The}} LAT photon flux light curve (red) can be well fitted by a power-law. A signature of observation with GeV emission during \edit1{\added{the}} X-ray flare also can be found. And the bottom panel represents the individual LAT photons and their energies. \edit1{\added{The open circles represent those photons with a $\geq0.9$ probability of being associated with the burst, the red solid circles indicate the GeV photons ($\geq 0.9$).}} The photon index obtained by likelihood analysis is \edit1{\replaced{$\Gamma\sim-2.3$}{$\Gamma_{LAT} \sim -2.3$}}.}\explain{In the original version, for this picture, we did not compute the probability of those photons of being associated with the burst. In this version, we did. For the LAT analysis, we adopted the energy range from 100 MeV to 5 GeV in the first revision instead of the energy range from 100 MeV to 100 GeV in the original version. But we found that the fitting result of photon flux light curve shares a common power-law index ($1.54\pm0.02$) with the original version, and the photon index obtained by likelihood analysis is similar ($\sim -2.3$). To distinguish from the Lorentz factor $\Gamma$ in the discussion, we replaced the $\Gamma$ with $\Gamma_{LAT}$.} \label{fig:lightcurve_all}
\end{figure}

The results of our analysis that \edit1{\added{the}} temporal profile of the emission from GRB 180720B varies with energy from 10 keV to \edit1{\replaced{100 GeV}{5 GeV}} are shown in Figure \ref{fig:lightcurve_all}. It represents the whole \edit1{\replaced{lightcurves}{light curves}} from \edit1{\added{the}} prompt to afterglow phase on the top. While, the photons with energy of $\gtrsim$ 100 MeV in the $Fermi$-LAT data are presented \edit1{\replaced{in}{at}} the bottom of Figure \ref{fig:lightcurve_all}. The fact that \edit1{\added{the}} highest energy photons with energy of GeV \edit1{\replaced{was}{were}} observed during \edit1{\added{the}} X-ray flare has emerged.

\begin{figure}
\plotone{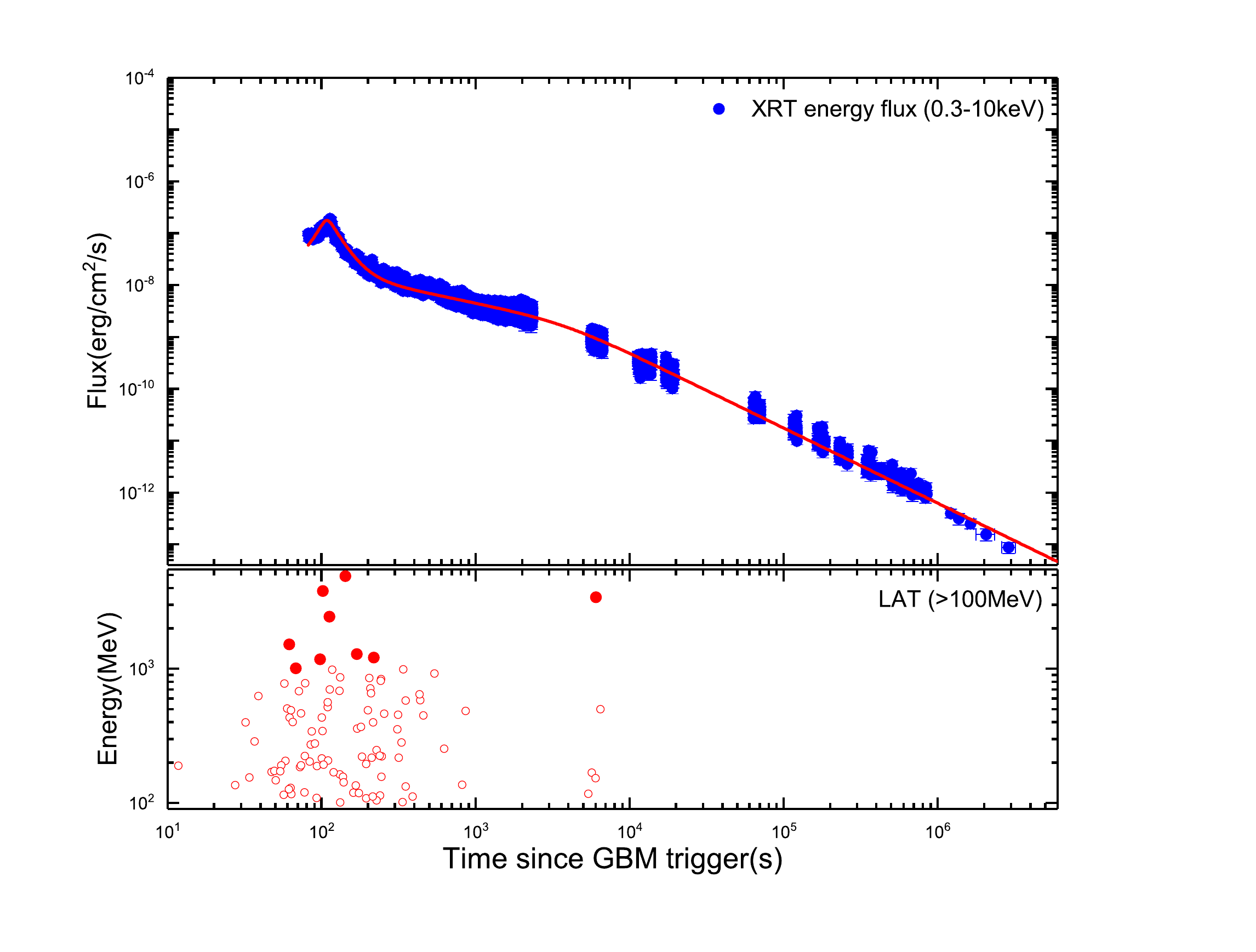}
\caption{Temporal characteristics. The upper panel shows \edit1{\added{the}} X-ray afterglow light curve of GRB 180720B fitted well by a double smoothly broken \edit1{\replaced{powerlaw}{power law}}. And the bottom panel represents the individual LAT photons and their energies. \edit1{\added{The open circles represent those photons with a $\geq0.9$ probability of being associated with the burst, the red solid circles indicate the GeV photons ($\geq 0.9$).}} A signature of observation with GeV emission during \edit1{\added{the}} X-ray flare also can be found.}\explain{In the original version, for this picture, we did not compute the probability of those photons of being associated with the burst. In this version, we did.} \label{fig:lightcurve_xrt}
\end{figure}

The GeV photons were detected at about 60 s after \edit1{\added{the}} trigger.
It's not disappeared until the highest energy photon (5 GeV, $\sim 142$ s) arise for this phenomenon. It remained that many MeV photons were observed in \edit1{\added{the}} afterglow. And the LAT photon flux light curve can be well fitted by a power-law read as: 
\begin{equation}
F_{1}=F_{01}t^{-\alpha}
\end{equation} (where $\alpha$ is the decay indice)
in Logarithmic Timescale with the best fit temporal index is $1.54\pm0.02$, which is similar to other $Fermi$-LAT bursts \citep{2013ApJS..209...11A}. One can find that the prompt gamma-ray emission detected by $Swift$/BAT is consistent with $Fermi$/GBM observation from Figure \ref{fig:lightcurve_all}. There is a very bright flare in \edit1{\added{the}} X-ray band while they were fading to a weaker level for GBM and BAT light curves. Without considering the fluctuation of the flare, it can be fitted well by a smoothly broken \edit1{\replaced{powerlaw}{power law}}, which is read as:
\begin{equation}
F=F_{02}[(\frac{t}{t_{b}})^{\omega\alpha_{1}}+(\frac{t}{t_{b}})^{\omega\alpha_{2}}]^{-1/\omega}
\end{equation}
where $\omega$ measures the sharpness of the peak. Then we get $\alpha_{X,1}=5.33\pm0.06$, $\alpha_{X,2}=-5.96\pm0.28$, $t_{b}=108s$, and $\omega=3$. The rapid increase and decrease of the flux imply that it would be the activity of \edit1{\added{the}} central engine in this burst. It means that it comes from the internal for this flare. Moreover, the X-ray afterglow light curve was fitted well by a double smoothly broken \edit1{\replaced{powerlaw}{power law}} (see Figure \ref{fig:lightcurve_xrt}).

On the other hand, to our surprise, the GeV observations of this burst are weaker than other LAT-bursts such as GRB 940217 \citep{1994Natur.372..652H}, GRB 130427A \citep{2013GCN.14471....1Z}, GRB 131231A \citep{2013GCN.15640....1S} and GRB 160625B \citep{2016GCN.19586....1D}, but, there are six photons with energy of GeV, a 1.2 GeV photon at $T_{0}+97.8 s$, a 3.8 GeV photon at $T_{0}+102 s$, a 2.45 GeV photon at $T_{0}+112 s$, a 5 GeV photon at $T_{0}+142 s$, a 1.3 GeV photon at $T_{0}+169 s$, a 1.2 GeV photon at $T_{0}+218 s$, after \edit1{\added{the}} trigger observed by LAT during the first X-ray flare while the lower energy emission \edit1{\replaced{are}{is}} fading. This implies the fact that the GeV flare \edit1{\replaced{arise}{arises}} at the same time the X-ray flare \edit1{\replaced{appear}{appears}}. Moreover, it also means that the GeV emission associated with \edit1{\added{the}} X-ray flare.

\subsubsection{Spectral Analysis} \label{subsubsec:3.2.2}

\begin{figure}
\plotone{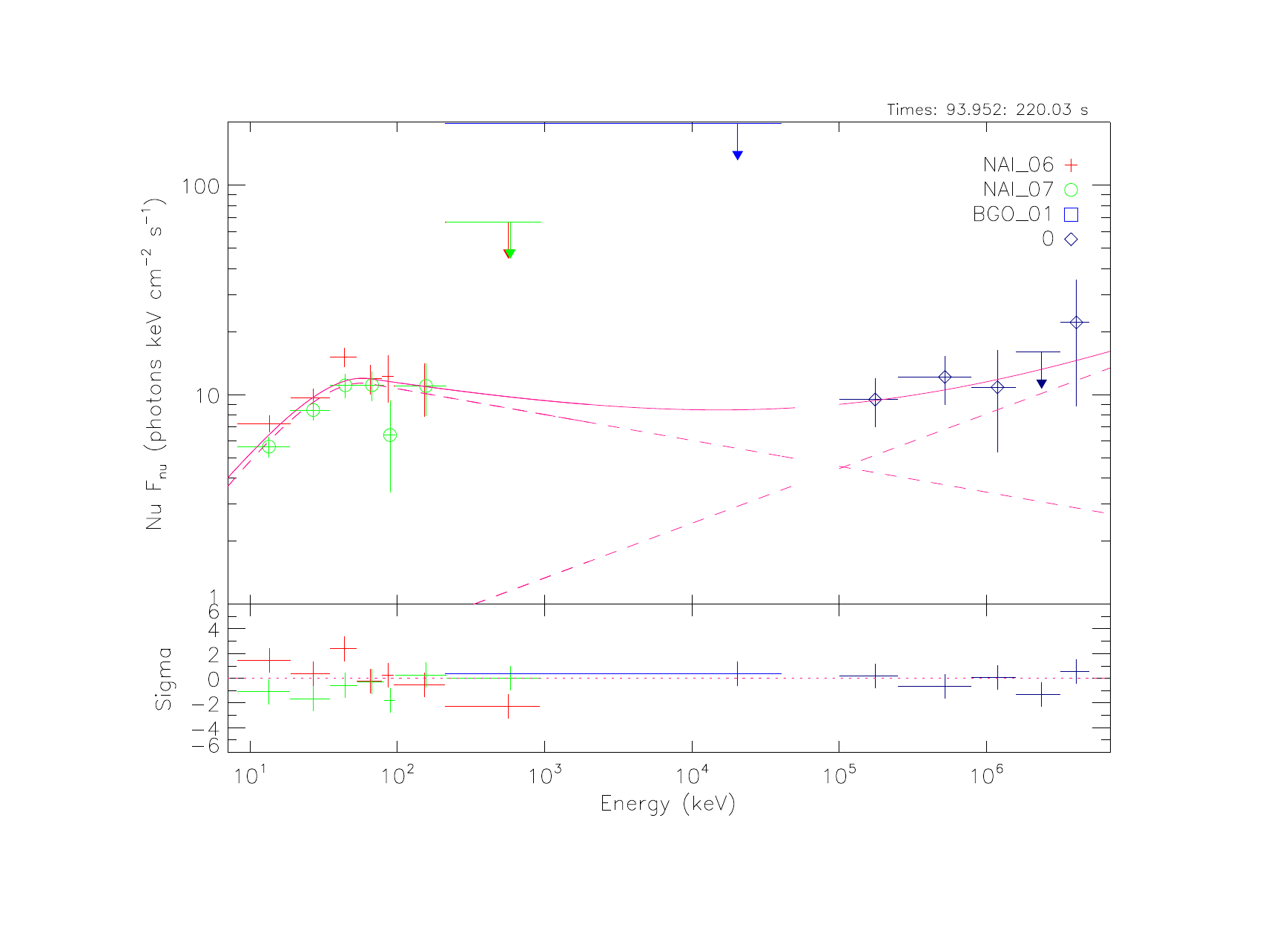}
\caption{Spectral characteristics. The joint GBM and LAT spectrum  from 94 s to 220 s. The \edit1{\replaced{blue}{deep pink}} solid line shows the best-fit model of the spectrum described in Section \ref{subsubsec:3.2.2}, i.e., the Band function with the \edit1{\replaced{superpositon}{superposition}} of \edit1{\replaced{powerlaw}{power law}}. In \edit1{\added{a}} lower energy range, about 8 keV to 200 keV, it's \edit1{\replaced{dominanted}{dominated}} by Band function for the spectrum. But in \edit1{\added{the}} high energy range (100 MeV to GeV), the \edit1{\replaced{powerlaw}{power law}} function is the most important instead of Band function.}\explain{In the original version, we selected the detectors $n_{6}$ and LAT. In this version, we selected the detectors $n_{6}$, $n_{7}$, $b_{1}$ and LAT even though there is no detection in BGO ($b_{1}$) energy range, the two upper limits were given.}
\label{fig:spectrum2}
\end{figure}

The results of our analysis of spectral energy distributions in \edit1{\added{the}} afterglow phase from lower energy to GeV range are shown in Figure \ref{fig:spectrum2} according to the photon statistics permission. The $\nu F_{\nu}$ spectrum (from about 94 s to 220 s after \edit1{\added{the}} trigger) by using the data select combined with \edit1{\replaced{NaI ($n_{6}$)}{NaI ($n_{6}$, $n_{7}$), BGO ($b_{1}$)}} and LAT is well fitted by Band function \citep{1993ApJ...413..281B} with the superposition of \edit1{\replaced{powerlaw}{power law}} which is extended to high energy (\edit1{\replaced{$\chi^{2}/dof=0.95$}{$\chi^{2}/dof=1.10$}}). \edit1{\replaced{For the first function, $E_{peak}\sim47.3\pm18.2$ keV, $\alpha_{1}\sim -0.916\pm0.620$, $\beta\sim -2.165\pm0.406$, and $\alpha_{2}\sim -1.738\pm0.516$ for the second function.}{For the first function, $E_{peak}\sim57.21\pm18.40$ keV, $\alpha_{1}\sim -1.07\pm0.46$, $\beta\sim -2.12\pm0.33$, and $\alpha_{2}\sim -1.74\pm0.61$ for the second function.}} \explain{We added the detectors $n_{7}$ and $b_{1}$ in this revision. So, the fitting results have varied.} At the same time, the neutral hydrogen density of the Milky Way in the burst direction is $N_{H}=3.92\times10^{20}cm^{-2}$. With the neutral hydrogen \edit1{\replaced{absorbtion}{absorption}} of the GRB host galaxy is taken into account, we found that a single power-law function is adequate to fit the X-ray time-averaged spectrum with the photon index $\Gamma_{X}=-1.76\pm0.03$ ($\chi^{2}/dof=1.11$). To our excitement, the Band function with the superposition of \edit1{\replaced{powerlaw}{power law}} can be extrapolated to the X-ray range from $Fermi$ energy range. In a words, the joint XRT, GBM and LAT spectrum can be fitted well by a Band function with the superposition of \edit1{\replaced{powerlaw}{power law}} which is \edit1{\replaced{dominanted}{dominated}} in high energy emission. And we obtain the photon index \edit1{\replaced{$\Gamma\sim-2.3$}{$\Gamma_{LAT} \sim -2.3$}}\explain{To distinguish from the Lorentz factor $\Gamma$ in the discussion, we replaced the $\Gamma$ with $\Gamma_{LAT}$.} (shown in Figure \ref{fig:lightcurve_all}) by making likelihood analysis using all 100 MeV to \edit1{\replaced{100}{5}} GeV photons, which is consistent with the indices of other LAT \edit1{\replaced{burst}{bursts}} \citep{2013ApJS..209...11A}. In other words, the spectral index $\beta_{LAT} \sim 1.3$ is available for $F_{\nu}\propto \nu^{-\beta}t^{-\alpha}$ because of $\beta_{LAT}\sim -\Gamma_{EXT}-1$ (EXT is the interval between the end of GBM detected and LAT-detected emission), where is read as: \edit1{\replaced{$\Gamma$}{$\Gamma_{LAT}$}} \explain{To distinguish from the Lorentz factor $\Gamma$ in the discussion, we replaced the $\Gamma$ with $\Gamma_{LAT}$.} \citep{2013ApJS..209...11A}.

We present the photons with energy of $\gtrsim$ 100 MeV in the $Fermi$-LAT data in the bottom panel of Figure \ref{fig:lightcurve_all} and Figure \ref{fig:lightcurve_xrt}. As shown in Figure \ref{fig:lightcurve_all}, there are some photons with energy of $\gtrsim$ 100 MeV while the low energy emission is fading to a weaker level. There are six GeV photons during the first X-ray flare. \edit1{\replaced{It's not impossible that the data throw}{The data may throw}} out a challenge to the theoretical model at any moment. It's attaching us to search for the mechanism of those monstrous photons in this burst, more details will be discussed later.

\subsubsection{Comparison of the very high energy emission of GRB 180720B with GRB 100728A, GRB 131231A and GRB 130427A} \label{subsubsec:Comparison}
There are many LAT-bursts since the launch of $Fermi$ from 2008. The photons with energy of $\gtrsim$ 100 MeV usually are detected  in \edit1{\added{the}} prompt or afterglow phase for gamma-ray bursts. And it's reported that the GeV photons were detected (include tens of GeV) sometimes such as GRB 100728A \citep{2011ApJ...734L..27A}, GRB 130427A \citep{2013GCN.14471....1Z} and GRB 131231A \citep{2013GCN.15640....1S}. They are also very particular. 

The redshift $z=0.654$ of the burst is similar to GRB 131231A with the redshift of $z=0.643$, which is a factor of two larger than GRB 130427A. While the redshift of GRB 100728A is a factor of two larger than the GRBs 180720B and 131231A. Photons with \edit1{\added{the}} energy of tens of GeV were found in GRBs 131231A, 130427A and GRB 100728A  while in GRB 180720B just the GeV photons within 10 GeV were collected. We note that the isotropic energy $E_{\gamma,iso}\sim 10^{53} ergs$ of GRB 180720B is similar to GRB 131231A in \edit1{\added{the}} prompt phase. This value is a factor of 10 smaller than GRB 130427A and GRB 100728A. But, there are something in common with GRB 100728A for this burst. Some GeV photons are observed by LAT during \edit1{\added{the}} X-ray flare both in GRB 180720B and GRB 100728A\citep{2013ApJ...772..152W}. All four have similar photon index with \edit1{\replaced{$\Gamma \sim -2$}{$\Gamma_{LAT} \sim -2$}} \explain{To distinguish from the Lorentz factor $Gamma$ in the discussion, we replaced the $\Gamma$ with $\Gamma_{LAT}$} (In general, it is $\sim -2$ for LAT-bursts.). 

With these similarities, we infer that they are produced with \edit1{\added{a}} similar physical mechanism to the GeV photons (or those photons with energy of 100 MeV to GeV) in GRBs 180720B, 131231A, 130427A and 100728A as it has been proved that the high energy component is produced by synchrotron self-Compton emission in refreshed shock originated from the \edit1{\replaced{reatctivation}{reactivation}} of \edit1{\added{the}} central engine in GRB 131231A \citep{2014ApJ...787L...6L}, GRB 130427A \citep{2014Sci...343...42A} and GRB 100728A \citep{2013ApJ...772..152W}. We expect that there are new discoveries, and more details will be discussed later.

\section{Discussion} \label{sec:Discussion}
\subsection{Origin of the Prompt Spectral Evolution Characteristics for GRB 180720B} \label{subsec:4.1}
For the features of GRB 180720B in \edit1{\added{the}} prompt, they can be summarized as: (i) the prompt emission is a multipulse structure; (ii) the `flux tracking' \edit1{\replaced{behaviour}{behavior}} emerged both for $E_{p}$ and $\alpha$; (iii) four parameter relations, $E_{p}-F$, $\alpha-F$, $E_{p}-\alpha$ and $E_{p}-L_{\gamma,iso}$, exhibit \edit1{\replaced{a strong positive correlation}{the strong positive correlations}} during the \edit1{\added{prompt}} interval; (iv) the value of lower energy photon index $\alpha$ does not exceed the synchrotron limits\edit1{\replaced{.}{;}} \edit1{\added{(v) The joint GBM and LAT (include LAT-LLE) time-averaged $\nu F_{\nu}$ spectrum from 11 s to 55 s can be fitted with a single Band function.}} In the following, we will discuss the origin of the prompt emission in GRB 180720B through the `flux-tracking' \edit1{\replaced{behaviour}{behavior}} for the $E_{p}$ and $\alpha$ of spectral evolution within the frameworks of the synchrotron and photosphere model.

The relation of $E_{p}\propto L^{1/2}\gamma_{e,ch}^{2}R^{-1}(1+z)^{-1}$ can be derived from \citet{2002ApJ...581.1236Z} in the synchrotron model, where $L$ is the `wind' luminosity of ejecta, $\gamma_{e,ch}$ is the typical electron Lorentz factor of emission region, $R$ is the emission radius, and $z$ is the redshift of the burst. Then, it is possible that the tracking \edit1{\replaced{behaviour}{behavior}} emerged since the relation of $E_{p}\propto L^{1/2}$. On the other hand, \citet{2018ApJ...869..100U} \edit1{\replaced{potinted}{pointed}} out that the `flux-tracking' \edit1{\replaced{behaviour}{behavior}} could be reproduced successfully within the synchrotron radiation model. In this model, for the $\alpha$ evolution with `flux-tracking' pattern, the $\alpha$ is increasing/decreasing while the flux is increasing/decreasing. As said in \citet{2019arXiv190104925L} for GRB 131231A, it could be attributed to the fact that the electron distribution is getting harder if it is the synchrotron origin for the hardening $\alpha$ before the first highest peak. Both the decaying magnetic field in the emission region \citep{2014ApJ...785..112D} and the synchrotron self-Compton (SSC) cooling of electrons \citep{2018ApJS..234....3G} can make the electron spectrum hardening. The increase of Lorentz factor $\gamma_{e,ch}$ of emitting electrons when the $R$ is larger and the decay of magnetic field will make the electron spectrum to be hard. And the ratio of the radiation energy to the magnetic energy is rising since the flux is increasing, then, the SSC cooling of electrons is more \edit1{\replaced{signficant}{significant}} so that the $\alpha$ to be harder.

Similarly, the relation of $E_{p}\propto L^{-5/12}r_{0}^{1/6}\Gamma^{8/3}$ for the $R_{ph}>R_{s}$ and the relation of $E_{p}\propto L^{1/4}r_{0}^{-1/2}$ for the $R_{ph}<R_{s}$ can be derived for photosphere model, where $r_{0}$ is the initial acceleration radius, $\Gamma$ is the bulk Lorentz factor, $R_{ph}$ is the radius of photosphere, and $R_{s}$ is the saturation radius. The anti-correlation is found between $E_{p}$ and $L$. \citet{2014ApJ...785..112D} pointed out that the observed `hard-to-soft' and `flux-tracking' \edit1{\replaced{behaviour}{behavior}} are both not easy to be reproduced in this model when they satisfy the relation of $R_{ph} > R_{s}$ between the radius of photosphere and saturation radius unless \edit1{\replaced{acertain}{a certain}} dependence between $\Gamma$ and $L$ is exist. However, the two observed $E_{p}$ evolution patterns of `hard-to-soft' and `flux-tracking' can be reproduced when the $R_{ph} <R_{s}$ since the relation of $R_{ph}\propto L$ is certain (Meng et al. 2019, which is told in preperation in \citet{2019arXiv190104925L}). On the other hand, in this model, the `hard-to-soft' pattern is predicted for the $\alpha$ evolution, which is different from complexity of the $E_{p}$ evolution.

For GRB 180720B, both the $E_{p}$ and $\alpha$ track the flux tightly \edit1{\added{in the prompt}}. In consideration of these prompt spectral evolutions, the interpretation that synchrotron origin can account for the coexist of the `flux-tracking' \edit1{\replaced{behaviours}{behaviors}} for the $E_{p}$ and $\alpha$, but the photosphere origin is invalid. And another important evidence that can support the interpretation of synchrotron origin is that the values of $\alpha$ during the \edit1{\added{prompt}} interval do not exceed the synchrotron limits by performing the detailed time-resolved spectral analysis. \edit1{\added{Besides, the most important is that the joint GBM and LAT (include LAT-LLE) time-averaged $\nu F_{\nu}$ spectrum from 11 s to 55 s in the prompt phase can be fitted well by a Band function without the additional power-law component extend to the LAT emission (see Section \ref{subsubsec:3.1.1}), which indicates that the low energy emission and high energy emission in the prompt phase from the burst share the common origin. Then, there is no doubt that the synchrotron origin can account for the mechanism of the prompt emission in this burst from lower energy to 1 GeV emission.}}

\subsection{Origin of the High Energy Emission during the X-ray Flare for GRB 180720B}\label{subsec:4.2}
\edit1{\deleted{GRB 180720B was detected by $Fermi$ (GBM and LAT), simultaneously, it also was detected by $Swift$ (BAT and XRT) and all of these data are available. For GRB 180720B, the fluence at  10 keV to 1000 keV measured with the GBM during the $T_{90}$ interval is $2.985 \pm 0.001 \times10^{-4} erg/cm^{2}$ \citep{2018GCN.22981....1R}, hence, the isotropic $\gamma$-ray energy is $E_{\gamma,iso}=3.3\times10^{53} ergs$ calculated by Equation (1) in \citet{2001AJ....121.2879B}, using a flat $\Lambda$CDM cosmology with reduced Hubble constant $H_{0}=71 (km/s)/Mpc$, dark matter density $\Omega_{M}=0.27$; and such a value represent a luminosity distance of 3916.6 Mpc for $z=0.654$. On the other hand, it implies a isotropic luminosity with value of $6\times10^{51}erg/s$ on average over the $T_{90}$ interval. }}\explain{We deleted this paragraph in this revision because we thought that these descriptions are essentially repetitive compared with the above such as Section \ref{sec:GRB 180720B} and the right panel in Figure \ref{fig:correlation2} in Section \ref{subsubsec:3.1.3}.}

Here, we expect that the high energy emission can be used to constrain the afterglow model. A spectrum as $F_{\nu}\propto \nu^{-1.3} t^{-1.54\pm0.02}$ (see Section \ref{subsubsec:3.2.1}, \ref{subsubsec:3.2.2}) from 100 MeV to \edit1{\replaced{100}{5}} GeV \edit1{\deleted{(in fact, the highest is 5 GeV)}} in LAT-bursts can be roughly accounted not only by synchrotron radiation\edit1{\deleted,} but also can be interpreted by synchrotron self-Compton radiation. It can be accounted \edit1{\added{for}} by synchrotron radiation if the injected electrons \edit1{\replaced{has index with value}{have an index with a value}} of $p\sim 2.6$ and it is above both the cooling frequency ($\nu_{c}$) and typical frequency ($\nu_{m}$) for the band selected by us \citep{2004IJMPA..19.2385Z}. Another case is synchrotron self-Compton radiation with \edit1{\added{an}} index of $p\sim 2.6$ and the band is above both the $\nu_{c}$ and $\nu_{m}$ without \edit1{\added{the}} evolution of Y named Compton parameter \citep{2007ChJAA...7..509W,2008FrPhC...3..306F}. One pointed out that it is likely to be the fast-cooling part for LAT-detected $\gtrsim$ 100 MeV emission \citep{1998ApJ...497L..17S,2002ApJ...568..820G}. Whereas, it's also available to use the interpretation by fast-cooling for both the first and second case. \edit1{\replaced{We will adrress the question: how to identify the mechanism of the high energy emission in this burst}{In the next 4 paragraphs, we will address the question: how to identify the mechanism of the high energy emission during the X-ray flare in the afterglow for this burst}}, synchrotron radiation or synchrotron self-Compton radiation?

Firstly, for the GeV emission, the two conditions are required to allow the GeV emission to be observed. One is that the source has to be optically thin for pair production as said in \citet{2011ApJ...734L..27A}, the lower limit on the Lorentz factor is 
\begin{equation}
\Gamma\geq\Gamma_{\gamma\gamma}\approx30E_{GeV}^{1/6}t_{v}^{-1/6}D_{28}^{1/3}(\frac{1+z}{2})^{1/3} 
\end{equation} (Equation (1) in \citet{2011ApJ...734L..27A}) when the value of 2 was thought as photon index value from \citet{2001ApJ...555..540L}. Then the $\Gamma$ will change while the $t_{v}$ named the shortest time scale is variable. If the X-ray flare is related to the activity of \edit1{\added{the}} central engine, one derives $\Gamma_{\gamma\gamma}\sim95E_{GeV}^{1/6}$ by \edit1{\replaced{asuming}{assuming}} $t_{v}=10^{-3}$ s for this burst. Based on the above assumption, $\Gamma\approx124.22$ is required to allow the 5 GeV photon to be observed at 142 s after the burst. And another condition is derived by requiring that the $\Gamma$ is large enough to make the blast wave accelerate electrons produce photons of energy $E_{GeV}$ through synchrotron radiation:
\begin{equation}
\Gamma>60(\frac{1+z}{2})E_{GeV}
\end{equation}
(Equation (3) in \citet{2011ApJ...734L..27A}) without the consideration of Y (named Compton parameter) evolution. We can derive the constrain of Lorentz factor in \edit1{\added{the}} emitting region, i.e., $\Gamma>248$ from the \edit1{\replaced{two condition}{two conditions}}. Whereas, \edit1{\deleted{it is impossible for this burst due to}} the low critical initial Lorentz factor $\Gamma_{c}\sim185$ \edit1{\added{was}} derived from the Equation (1) in \citet{2003ApJ...595..950Z} \edit1{\added{for this burst}}. It makes the Lorenz factor at any moment in \edit1{\added{the}} afterglow is lower than this value \edit1{\replaced{because}{because of}} the relationship of $\Gamma\propto t^{-\frac{3}{8}}$. Based on the two conditions, we can point out that, simply but robustly, \edit1{\replaced{it is not possible that the photons with value of GeV are produced by synchrotron radiation.}{synchrotron radiation cannot produce the photons with the value of GeV.}} In fact, the GeV emission has been thought to arise from external inverse Compton (EIC) or synchrotron self-Compton (SSC) \citep{2008MNRAS.384.1483F}. And it's worth noting that, one pointed out that, it seems that they are produced by EIC instead of SSC for the GeV photons during the X-ray flare detected by $Fermi$/LAT \citep{2006MNRAS.370L..24F,2008MNRAS.384.1483F,2006ApJ...641L..89W,2012ApJ...753..178H}, while the GeV photons were observed during the first X-ray flare in this burst as said above. Here, \edit1{\replaced{we adopt another different explanation}{another different explanation will be useful}}, in which the detected GeV photons are produced by SSC in the refreshed shock from the reactivation of \edit1{\added{the}} central engine. \edit1{\deleted{The details are represented in below.}} As mentioned in \citet{2008MNRAS.384.1483F}, there are some differences between EIC and SSC for the GeV emission during the X-ray flare. They originate from the reactivation of \edit1{\added{the}} central engine both for EIC and SSC, but, the latter is produced by the interaction from photons and electrons both in shocks. And another one comes from the interaction between photons (those photons move to external shocks later) and hot electrons in the external shocks. In a word, the former is originated by later internal shock but it's produced by the refreshed shocks due to the encounter between \edit1{\added{the}} later shell and external shocks. Just because this, \edit1{\replaced{the more higher energy}{higher energy}} photons would arrive at \edit1{\added{a}} later time compared with the lower photons in the former, but, they arrive at the same time in the latter. In a word, it's expected that there are observations of \edit1{\added{the}} GeV flash \edit1{\replaced{associate}{associated}} with \edit1{\added{the}} X-ray flare both in \edit1{\replaced{foremer}{the former}} and latter, but we can identify them through comparing the time of arrival between GeV flash photons and X-ray flare photons. No doubt, in our burst, it's obvious that the maximal probability is synchrotron self-Compton for GRB 180720B.

Secondly, for the LAT emission, the $E_{k}$ with \edit1{\added{the}} value of $\sim 10^{57} ergs$ is necessary to produce flux $\sim 10^{-7} erg/cm^{2}/s$ at $\sim100$ s for synchrotron radiation. It seems unrealistic. There is no doubt that it should \edit1{\replaced{be gave up}{give up}} to give a reasonable interpretation using the thought that the LAT emission derived from \edit1{\replaced{shychrotron}{synchrotron}} radiation. While\edit1{\deleted{,}} the interpretation that the LAT emission \edit1{\replaced{arised}{arisen}} from synchrotron self-Compton radiation can be called up for this burst. In \edit1{\added{the}} synchrotron self-Compton radiation model, as summarized by \citet{2008FrPhC...3..306F}, it is possible to produce such a flux. From the Equations (52) and (53) in  \citet{2008FrPhC...3..306F}, through \edit1{\replaced{estemating}{estimating}} the value of flux with the equation:
\begin{equation}
\nu F_{\nu}=\nu(\frac{\nu}{\nu_{m}})^{-\frac{p}{2}}(\frac{\nu_{m}}{\nu_{c}})^{-\frac{1}{2}}F_{\nu,max}
\end{equation}
The flux $\sim 10^{-7} erg/cm^{2}/s$ is allowed during the interval using the two characteristic frequencies and the above equation in SSC. As described above, it is very reasonable to interpret the spectrum using synchrotron self-Compton radiation. The model $\beta \sim \frac{p}{2}$, $\alpha\sim\frac{9p-10}{8}$ without evolution of Y (Compton parameter) \citep{2007ChJAA...7..509W,2008FrPhC...3..306F} is consistent with both the spectrum data and the temporal \edit1{\replaced{behaviour}{behavior}} in the environment of \edit1{\replaced{instellar}{interstellar}} medium when electrons are in the fast cooling phase.

Thirdly, it is inadequate to fit the joint spectrum (XRT, GBM and LAT) as described in Section \ref{subsubsec:3.2.2} by using a single function. The superposed \edit1{\replaced{powerlaw}{power-law}} function has to be used to \edit1{\replaced{extended}{extend}} to the high energy emission of LAT instead of \edit1{\added{a}} single Band function. And it's found that it is important both for GeV energy range and sub-GeV energy range in LAT. In general, \edit1{\replaced{they are thought as the origin of synchrotron radiation for the lower energy emission which is dominated by Band}{the low energy emission which is dominated by Band function is thought as the origin of synchrotron radiation}}. So, \edit1{\added{to interpret the additional power-law component,}} the synchrotron self-Compton radiation must be called for.

\edit1{\replaced{All of the above three points give a clear interpretation that the LAT emission during X-ray flare was originated from another mechanism instead of traditional synchrotron radiation. Therefore, one cannot refuse to accept the interpretation that synchrotron self-Compton radiation can account for this afterglow emission based on the above discussion. At last, we can regard them as the internal origin if the seed photons (X-ray photons) with lower energy were produced due to the reactivation of central engine.}{All of the three points in the past 3 paragraphs gives a clear interpretation that those high energy photons during the X-ray flare were originated from another mechanism instead of the traditional synchrotron radiation, which means one can accept the interpretation that synchrotron self-Compton radiation can account for this type of high energy afterglow emission. Moreover, if the seed photons with lower energy were produced due to the reactivation of the internal central engine, we can regard them as the internal origin for the high energy photons during the X-ray flare. In a word, they originated from the internal and produced by synchrotron self-Compton  radiation for those high energy photons during the X-ray flare for GRB 180720B.}}

\section{Conclusion} \label{sec:5}
GRB 180720B is a long, very bright and peculiar burst with the multipulse structure \edit1{\deleted{consists of an initial continuedly multipeaked emission episode lasting for a dozen seconds ($\sim$ 8-20 s), a sharp pulse with lower amplitude at about 30 seconds after trigger, and another sharp pulse with lowest amplitude at about 50 s, which fades to a weaker level after $\sim$ 200 s}}. It can be \edit1{\replaced{represent for}{representative of}} the multipulse bursts. \edit1{\added{In this work, we performed the detailed temporal characteristics analysis and spectral analysis both for the prompt phase and afterglow phase in this burst. There are some interesting features in our study:
\begin{enumerate}
\item The `flux-tracking' pattern exhibited both for $E_{p}$ and $\alpha$ in the prompt phase.
\item There are four strong positive correlations such as $E_{p}-F$, $\alpha-F$, $E_{p}-\alpha$ and $E_{p}-L_{\gamma,iso}$ during the prompt interval.
\item The value of lower energy photon index $\alpha$ does not exceed the synchrotron limits in the prompt phase.
\item A single Band function is adequate to fit the joint GBM and LAT (include LAT-LLE) spectrum from 11 s to 55 s after the GBM trigger.
\item There are six GeV photons were detected during the X-ray flare in the afterglow.
\item The joint GBM and LAT spectrum can be fitted well by the Band function with the superposition of power-law extends to the LAT range when the six GeV photons were detected.
\end{enumerate} }}

 \edit1{\replaced{The evolution of $\alpha$ and $E_{p}$ with `flux-tracking' behaviour imply that it is synchrotron origin for prompt emission.}{The fact that the Band function can fit the joint time-averaged $\nu F_{\nu}$ spectrum without the additional power-law component extend to the LAT emission and the $E_{p}$, $\alpha$ exhibit the `flux-tracking' patterns implies that it is synchrotron origin in the prompt phase for GRB 180720B.}} And at last, we must accept the interpretation that the high energy emission during \edit1{\added{the}} X-ray flare is originated from synchrotron self-Compton radiation in light of all the \edit1{\replaced{evidences in Section\ref{sec:Discussion}}{evidence in Section \ref{subsec:4.2}}}. 

\acknowledgments

We thank all the people for helpful suggestions. We acknowledge the use of the public data from the Fermi data archives. And this work made use of data supplied by the UK Swift Science Data Centre at the University of Leicester. This work is supported by the National Natural Science Foundation of China (grant No.11673006), the Guangxi Science Foundation (grant Nos. 2016GXNSFFA380006, 2017AD22006,  2018GXNSFDA281033), the One-Hundred-Talents Program of Guangxi colleges, and High level innovation team and outstanding scholar program in Guangxi colleges.



\listofchanges

\end{document}